\documentclass{article}

\PassOptionsToPackage{numbers, compress}{natbib}

\usepackage[final]{neurips_2024_ml4ps}

\usepackage[utf8]{inputenc}       
\usepackage[T1]{fontenc}          
\usepackage{hyperref}             
\usepackage{url}                  
\usepackage{booktabs}             
\usepackage{amsfonts}             
\usepackage{nicefrac}             
\usepackage{microtype}            
\usepackage{xcolor}               

\usepackage{amsmath}              
\usepackage{graphicx}             
\usepackage{subcaption}           
\usepackage{soul} 
\usepackage{array}
\newcolumntype{M}[1]{>{\centering\arraybackslash}m{#1}}
\usepackage{wrapfig}
\usepackage{verbatim}
\usepackage{stmaryrd}

\graphicspath{{./figs/}}

\DeclareMathOperator*{\argmin}{arg\,min}

\title{Uncertainty quantification for fast reconstruction methods using augmented equivariant bootstrap: Application to radio interferometry}

\author{%
  Mostafa Cherif$^{1}$ \quad
  Tobías I.~Liaudat$^{1}$ \quad
  Jonathan Kern$^{1}$ \quad
  Christophe Kervazo$^{2}$ \quad
  Jérôme Bobin$^{1}$ \quad
  \\
  \\ 
  $^{1}$IRFU, CEA, Université Paris-Saclay, F-91191 Gif-sur-Yvette, France\\
  $^{2}$LTCI, Télécom Paris, Institut Polytechnique de Paris, 91120 Palaiseau, France\\
  \texttt{\{mostafa.cherif,tobias.liaudat,jonathan.kern,jerome.bobin\}@cea.fr} \\
  \texttt{christophe.kervazo@telecom-paris.fr}
}

\begin{document}

\maketitle

\begin{abstract}
The advent of next-generation radio interferometers like the Square Kilometer Array promises to revolutionise our radio astronomy observational capabilities. The unprecedented volume of data these devices generate requires fast and accurate image reconstruction algorithms to solve the ill-posed radio interferometric imaging problem. Most state-of-the-art reconstruction methods lack trustworthy and scalable uncertainty quantification, which is critical for the rigorous scientific interpretation of radio observations. We propose an unsupervised technique based on a conformalized version of a radio-augmented equivariant bootstrapping method, which allows us to quantify uncertainties for fast reconstruction methods. Noticeably, we rely on reconstructions from ultra-fast unrolled algorithms. The proposed method brings more reliable uncertainty estimations to our problem than existing alternatives.
\end{abstract}

\section{Introduction}

Radio interferometry \cite{Thompson2017} combines signals from multiple antennas to obtain images with very high resolutions. The next-generation radio-interferometry devices, such as the Square Kilometer Array (SKA) \cite{Dewdney2006}, will have the potential to unlock scientific discoveries thanks to their unprecedented angular resolution and sensitivity. However, this comes with the computational challenge of processing the incoming data deluge \cite{scaife2020}. Interferometric data comprises an incomplete coverage of the Fourier domain (uv-space), which, added to the observational noise, makes the reconstruction problem an ill-posed inverse problem, known as radio interferometric (RI) imaging. 

Most families of RI imaging methods, like CLEAN-based \cite{Hogbom1974, offringa2014, offringa2017}, sparsity-based \cite{mcewen2011, carrillo2012, carrillo2014, dabbech2018}, Bayesian approaches\cite{junklewitz2016, Arras2018, knollmuller2019, arras2021}, or learned iterative algorithms \cite{terris2022}, suffer from drawbacks such as bad reconstruction quality, due to constraints of the model considered, or long computing times, due to the iterative nature of their algorithm. Learned end-to-end algorithms use recent deep-learning techniques to offer ultra-fast and accurate image reconstructions \cite{Aghabiglou2024, mars2024, chen2022, Kern2024}. However, these models lack interpretability, making the attempt to quantify the reconstruction's uncertainty an arduous task.

It is essential to rigorously quantify our reconstruction's uncertainties to conduct scientific studies and make decisions based on these reconstructions. The high dimensionality of the RI data makes many standard uncertainty quantification (UQ) techniques, like MCMC sampling-based techniques \cite{Cai2017a} or conformal prediction \cite{Angelopoulos2021}, impractical or ineffective. Existing UQ methods rely on the Bayesian framework exploiting approximations \cite{pereyra2017} to avoid sampling \cite{Cai2017b, Liaudat2023}, but still rely on iterative optimisation techniques and do not exhibit tight bounds on the estimated uncertainty intervals. Alternative UQ methods rely on Gaussian priors \cite{junklewitz2016, Arras2018} and exploit variational inference \cite{knollmuller2019}. Recent methods, based on ensemble techniques, only quantify the model uncertainty \cite{Terris2023, Aghabiglou2024UQ}, failing to take into account the predominant source of uncertainties that stem from the forward operator's large null space due to the partial Fourier coverage. Recent methods based on score-based priors \cite{dia2023,sun2024,wu2024} obtain good results and provide UQ but are not adapted to the large-scale problems raised by SKA-type interferometers as they rely on sampling schemes.

We propose to use a fast and performant learned end-to-end reconstruction method, based on an unrolled architecture, equipped with a conformalized equivariant bootstrapping method adapted to the RI imaging problem. The equivariant bootstrap method is well suited to the RI imaging problem as, by selecting group actions adapted to the forward model's null space, we can reduce the bootstrapping estimation error, providing us with tight uncertainty bounds and excellent coverage plots. In addition, the subsequent conformalization procedure provides statistical guarantees to the estimated intervals.

\section{Radio-interferometric imaging problem}

We consider the simplified convolutional form of the observational model for the RI imaging problem, which writes
\begin{equation}
    \label{forward}
    y = M \ast x^{\star} + n ,
\end{equation}
where $y \in \mathbb{R}^{d \times d}$ represents our observations, i.e., the backprojected visibilities, $M \in \mathbb{R}^{d \times d}$ the point spread function (PSF) models the acquisition process including the incomplete uv-coverage (see Appendix~\ref{psf_uv_coverage} for examples), $x^{\star} \in \mathbb{R}^{d \times d}$ is the ground truth image, $\ast$ denotes convolution, and $n \in \mathbb{R}^{d \times d}$ models the observational and instrumental noise, which we suppose, for simplicity, to be white Gaussian additive noise with zero mean and a known standard deviation.

The most computationally expensive operation in RI imaging is applying the forward model due to the (de)gridding operations for the large number of visibilities observed. Therefore, drastically limiting the number of applications of the forward model is mandatory when considering large-scale applications like SKA. 
Most current RI imaging methods are iterative algorithms, e.g., sparsity-based and Plug-and-Play (PnP) \cite{venkatakrishnan2013}, that require many applications of the forward operator for convergence. 
In this work, we rely on an unrolled architecture \cite{Gregor2010, Monga2021}, where a fixed number of iterations of convex optimization algorithms \cite{Beck2009} are unfolded by representing all its operations as layers of a neural network. This choice is computationally efficient allowing it to achieve ultra-fast reconstructions while leveraging the knowledge of the forward operator and obtaining state-of-the-art reconstruction qualities \cite{Kern2024}. We use the EVIL-Deconv \cite{Kern2024} reconstruction method, where the authors build an unrolled architecture based on LISTA-CP \cite{Chen2018} using a DRUNET \cite{Zhang2020} for denoising. The EVIL-Deconv method uses the PSF as input and the model is trained on a set of PSFs covering various synthetic observational configurations. Therefore, it does not need to be retrained for a change in PSF.
A more detailed description of the method can be found in Appendix~\ref{Unrolling}. Despite the good performance of the method, the reconstruction lacks UQ.

\section{Uncertainty Quantification Methods}
\label{sec:uq_methods}

We start by summarizing the standard parametric bootstrapping procedure. In our forward model Eq.~\eqref{forward}, $y$ can be seen as a realization of the probabilistic model, $Y \sim P(M \ast x^{\star}) := P(\bar{M} x^{\star})$, where $P$ represents the noise distribution, and $\bar{M}$ a matrix representing the convolution by $M$. Starting from the reconstructed image $\hat{x}(y)$, we draw bootstrap measurements $y_i$ using the model in Eq.~\eqref{forward} and replacing $x^{\star}$ with $\hat{x}(y)$. Using the same reconstruction method, we then compute $x_i = \hat{x}(y_i)$, which we can compare to $\hat{x}(y)$. From the collection of $N$ bootstrap samples, $\{x_{i}\}_{i=1}^{N}$, we build confidence regions for $x^{\star}$. In this work, we consider the confidence region $\mathcal{C}_\alpha$ using $q_\alpha$ the top $\alpha$-quantile of the samples $\{ | x_i - \hat{x}(y) | \}_{i = 1}^N$, with $\mathcal{C}_\alpha = \{x : |x - \hat{x}(y)| < q_\alpha \}$. This conventional parametric bootstrap tends to underestimate the uncertainties in imaging problems where considerable uncertainties span from the forward operator's large null space\cite[\S 3]{Tachella2023}, as is the case in RI imaging.

Equivariant bootstrap \cite{Tachella2023} is based on recent developments in the equivariant imaging framework \cite{chen2021, Tachella2022} exploiting symmetries in the set of signals $\mathcal{X}$, with $x^{\star} \in \mathcal{X}$. Let us define $\mathcal{G}$ as a finite group acting on $\mathcal{X}$ with group actions represented by $T_{g_i} \in \mathcal{G}$, an invertible mapping. Assuming that $\mathcal{X}$ is $\mathcal{G}$-invariant, we can have access to multiple virtual forward operators, $\bar{M} T_{g_i}$, with possibly different null spaces if $\bar{M}$ is not $\mathcal{G}$-equivariant. The bootstrap method consist of drawing a random group action $T_{g_i} \in \mathcal{G}$, to then generate a bootstrap measurement $\Tilde{y}_i$ from $\Tilde{Y}_i \sim P(\bar{M} T_{g_i}\hat{x}(y))$. Later, we compute $\Tilde{x}_i = T_{g_i}^{-1}\hat{x}(\Tilde{y}_i)$. Finally, we can use the generated bootstrap samples $\{\Tilde{x}_i\}_{i = 1}^N$ to construct a confidence region $\mathcal{C}_\alpha$ as before, but using $\tilde{q}_\alpha$ the top $\alpha$-quantile of the samples $\{ | \Tilde{x}_i - \hat{x}(y) | \}_{i = 1}^N$. If $T_{g_i}$ is properly chosen based on the particularities of $\mathcal{X}$ and $\bar{M}$, the composition $\bar{M} T_{g_i}$ can have different null spaces than $\bar{M}$ helping to probe the variability of the estimator $\hat{x}(Y)$ and characterize its uncertainties with respect to $x^{\star}$.

\subsection{Conformalized Augmented Radio Bootstrap (CARB)}
\label{sec:CARB}

\begin{wrapfigure}[10]{R}{0.37\textwidth}
    \centering
    \vspace{-1.7\intextsep}
    \includegraphics[width=0.37\textwidth]{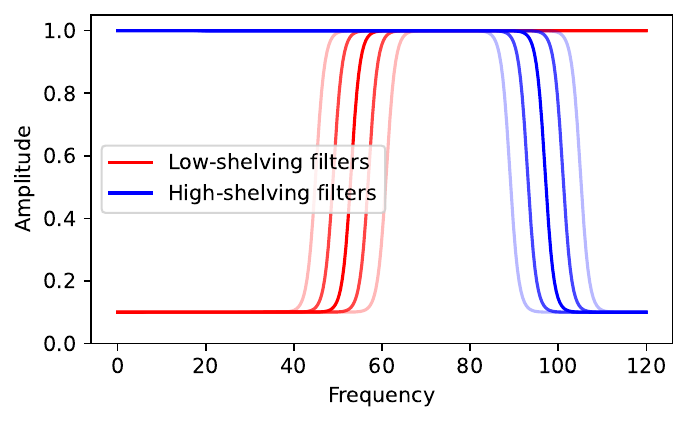}
    \vspace{-1.5\baselineskip}
\caption{1D radial visualization of shelving filters.}
\label{fig:shelving_filters}
\end{wrapfigure} 

We created a set of group actions calibrated to the RI imaging problem considered for our method, coined Conformalized Augmented Radio Bootstrap (CARB). First, we calibrate and include standard transformations as circular shift transformations not exceeding $2$ pixels, image flips over the horizontal and vertical axis, and rotations of $90$-degrees multiples, bringing the best results. Based on a convolution, the forward operator $\bar{M}$ is equivariant to translations. However, as the reconstruction operator is not, the composition $\bar{M} T_{g_i}(\hat{x}(y))$ spans a different subspace than $\bar{M} \hat{x}(y)$, making the translation actions useful. We then consider the group of invertible 2D filters in the specific form of low-shelving and high-shelving filters, as shown in Figure~\ref{fig:shelving_filters}. These filters attenuate low or high frequencies without cutting them off completely, assuring their inverses exist. The filters prove helpful and adapted to our model as they change the frequency response of the original filter represented by the PSF convolution. The resulting filter, a combination of the shelving filter and the PSF, will probably span a different subspace than the PSF alone. The change in the resulting filter helps to improve the characterisation of the errors from our estimation method, $\hat{x}(Y)$. The probability of applying each of the two filter transformations is $0.5$. The high and low drop-off frequencies are selected from two Gaussian distributions around frequencies we have calibrated for the RI imaging problem, with standard deviations of $5$. We add a constraint to ensure that the low drop-off frequency stays below the high drop-off frequency. The final group action used to generate the bootstrap samples in CARB is a random composition of the aforementioned transformations, where each transformation is applied with a given probability. This composition allows us to significantly expand the number of possible group actions, helping to estimate uncertainties better.

By constructing the confidence region with the $\alpha$-quantile, we could expect that for most images, only $100(1-\alpha)$\% of their pixels would have absolute residuals above the estimated uncertainty value. Once the confidence regions are estimated, we want to calibrate them to statistically guarantee that the uncertainty intervals around each pixel contain the true value with a particular user-chosen error rate, $\delta$, as is done in \cite{Angelopoulos2022} on heuristic-based uncertainties. Therefore, we include a subsequent conformalization procedure \cite{Angelopoulos2021} based on Risk-Controlling Prediction Sets (RCPS) \cite{Angelopoulos2022}, which we describe with more details in Appendix \ref{app:conf}. In practice, we fix the error rate $\delta$ to $0.1$ in our numerical experiments.

\section{Numerical experiments}
\label{sec:data}

We compare the proposed UQ method, CARB, with quantile regression \cite{Koenker1978}, conformalized quantile regression (CQR) \cite{Romano2019}, a standard parametric bootstrap, the equivariant bootstrap \cite{Tachella2023} (including rotations, flips, and up to $2$-pixel translations), and a version of CARB without conformalization. All the bootstrap-based methods use $500$ samples, which was selected based on a study shown in Appendix \ref{numberOfSamples}. To train our quantile regression method, we keep the same architecture of our unrolling network as the reconstruction, and replace the last layer with two learned neural networks. These networks are trained with the $5^\text{th}$ and $95^\text{th}$ quantile losses, which is a standard procedure \cite{Angelopoulos2021, Romano2019}. 

\paragraph{Data}{We use $64\times64$ patches from Hubble space telescope observations as ground truth images to keep the expensive numerical experiments in our computational budget. The simulated PSFs are based on the MeerKAT radio telescope \cite{Meerkat} antenna array. We train our unrolled architecture for an SNR range between $30$ to $60$dB. We use an SNR of $40$dB for the rest of the paper and suppose it is known. For all methods comparisons, $10000$ images were used to have statistically significant results.}

\subsection{Results} \label{results}

\begin{figure}

    \centering
    \vspace{-0.5\intextsep}
    \begin{subfigure}[t]{0.2\textwidth}
        \includegraphics[width=1\textwidth]{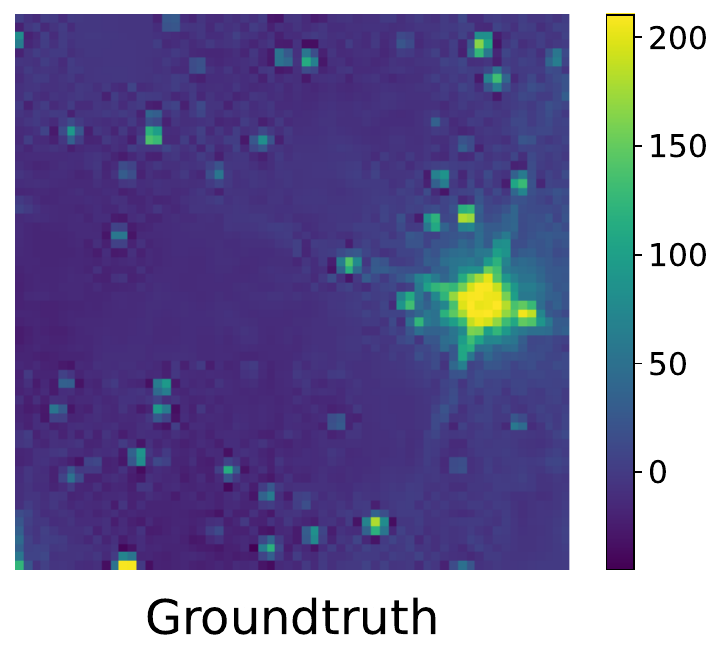}
    \end{subfigure}%
    \hspace{7pt}%
    \begin{subfigure}[t]{0.2\textwidth}
        \includegraphics[width=1\textwidth]{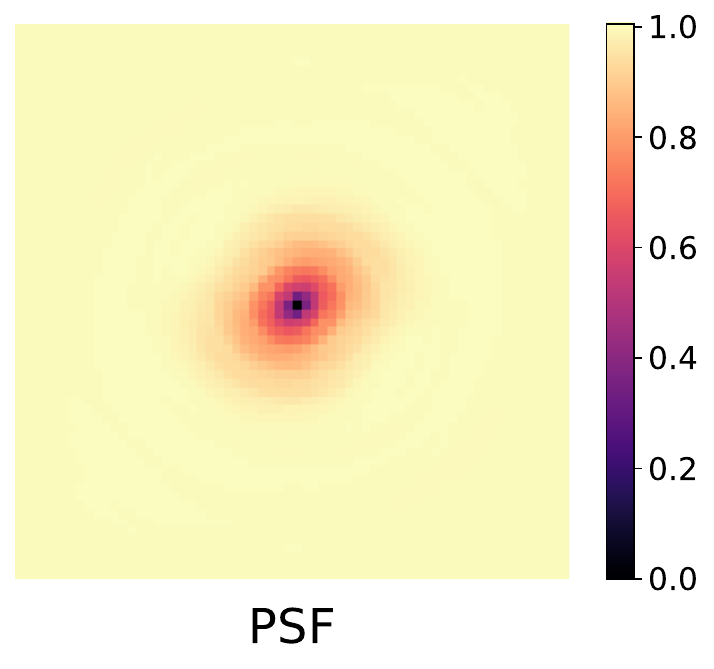}
    \end{subfigure}%
    \hspace{7pt}%
    \begin{subfigure}[t]{0.2\textwidth}
        \includegraphics[width=1\textwidth]{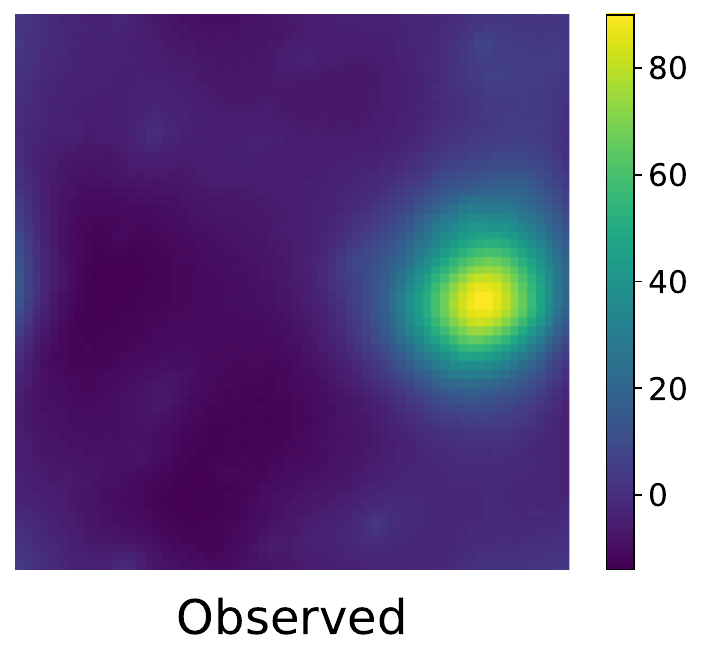}
    \end{subfigure}%
    \hspace{7pt}%
    \begin{subfigure}[t]{0.2\textwidth}
        \includegraphics[width=1\textwidth]{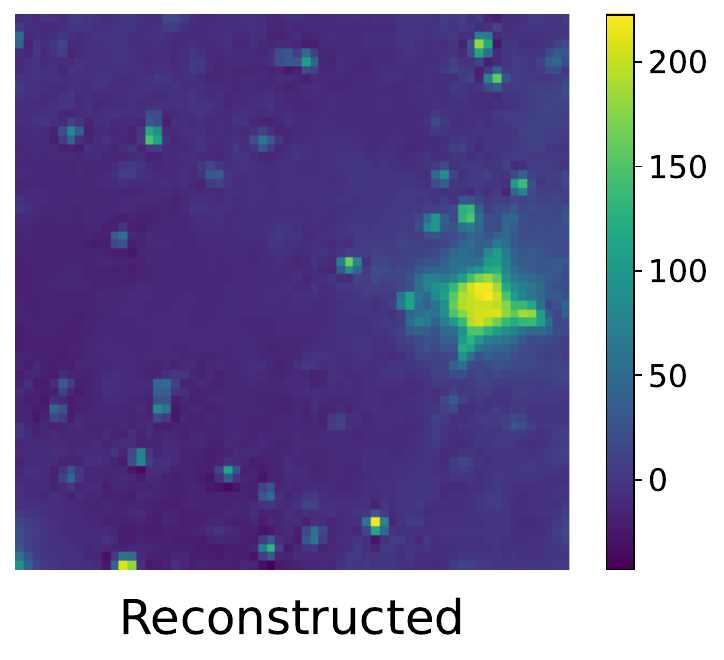}
    \end{subfigure}
    \vspace{-0.4\baselineskip}
    \caption{RI image reconstruction result for the unrolling algorithm. From left to right, the figures correspond to: the ground truth observation; the simulated point spread function (PSF); the observed image after the application of the PSF and the observational noise; and the reconstructed image from EVIL-Deconv.}
    \label{fig:unroll}
    \vspace{-1.\baselineskip}
\end{figure}  

We show an example of a RI image reconstruction result with the unrolled architecture in Figure~\ref{fig:unroll}, with additional results in Appendix~\ref{additional_results}. We then include a quantitative performance comparison with other state-of-the-art reconstruction methods in Appendix~\ref{app:table}.

In Figure~\ref{fig:error}, we compare the oracle, or ground truth, estimation error with each method's $90$-th quantile estimation to provide a pixel-wise UQ visualization of the reconstruction. We observe a high mismatch between the quantile regression and the oracle error. The confomalization procedure in the CQR tries to correct the mismatch by greatly inflating the quantiles, making them impractical for any interpretation. The equivariant bootstrap-based methods provide error maps with high correlation to the oracle error and tight error bars, which can be further calibrated with the conformalization procedure.

\begin{figure}[h]

    \vspace{-0.3\intextsep}
    \centering
    \begin{subfigure}[t]{0.19\textwidth}
        \includegraphics[width=1.05\textwidth]{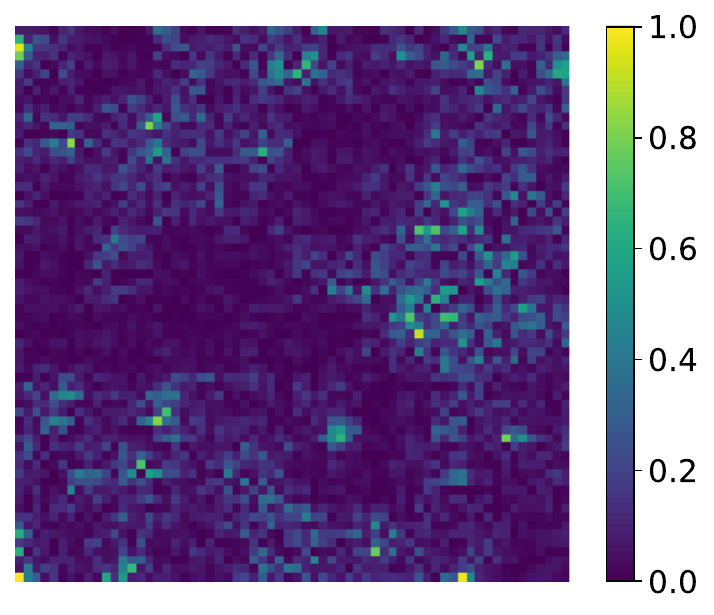}
        \vspace{-1.5\baselineskip}
        \caption{}
        \label{absres}
    \end{subfigure}%
    \hspace{10pt}%
    \begin{subfigure}[t]{0.21\textwidth}
        \includegraphics[width=1.0\textwidth]{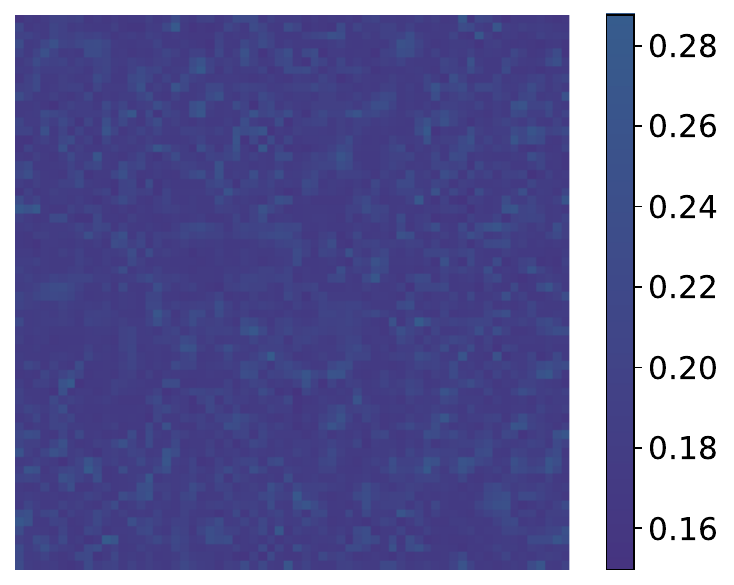}
        \vspace{-1.5\baselineskip}
        \caption{}
        \label{qr}
    \end{subfigure}%
    \hspace{10pt}%
    \begin{subfigure}[t]{0.21\textwidth}
        \includegraphics[width=1.0\textwidth]{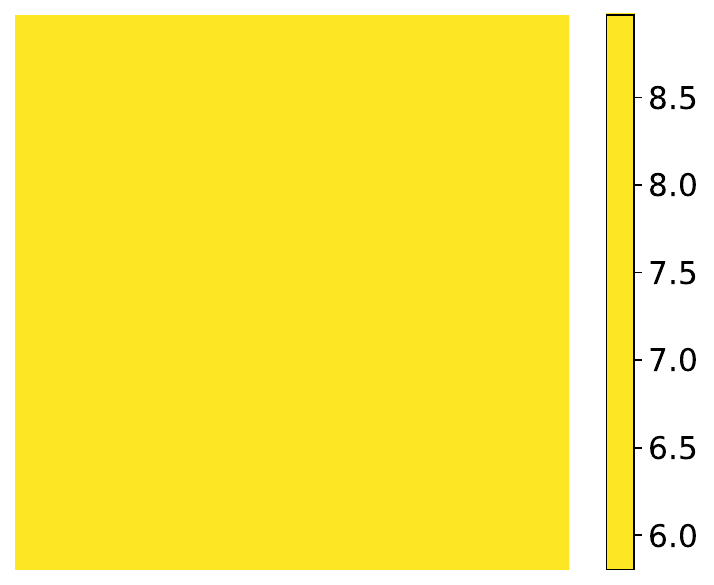}
        \vspace{-1.5\baselineskip}
        \caption{}
        \label{cqr}
    \end{subfigure}%
    \hspace{10pt}%
    \begin{subfigure}[t]{0.21\textwidth}
        \includegraphics[width=1.0\textwidth]{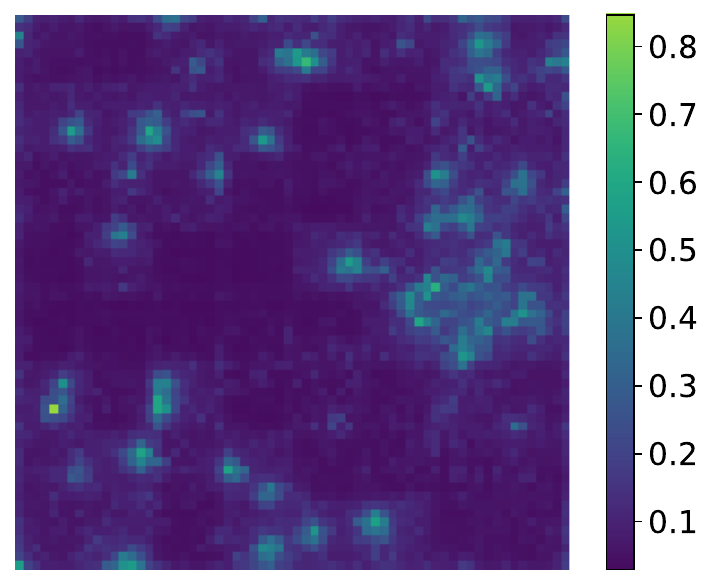}
        \vspace{-1.5\baselineskip}
        \caption{}
        \label{pb}
    \end{subfigure}%
    \\
    \hspace{10pt}%
    \begin{subfigure}[t]{0.21\textwidth}
        \includegraphics[width=1.0\textwidth]{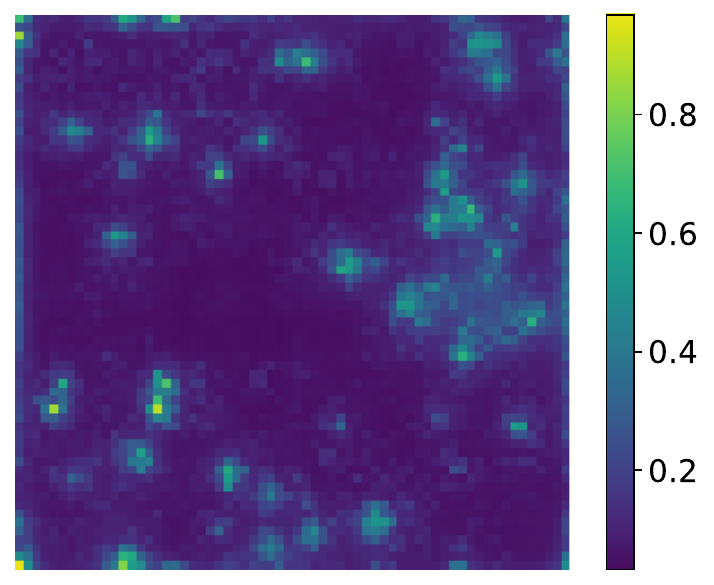}
        \vspace{-1.5\baselineskip}
        \caption{}
        \label{eb}
    \end{subfigure}%
    \hspace{10pt}%
    \begin{subfigure}[t]{0.21\textwidth}
        \includegraphics[width=1.0\textwidth]{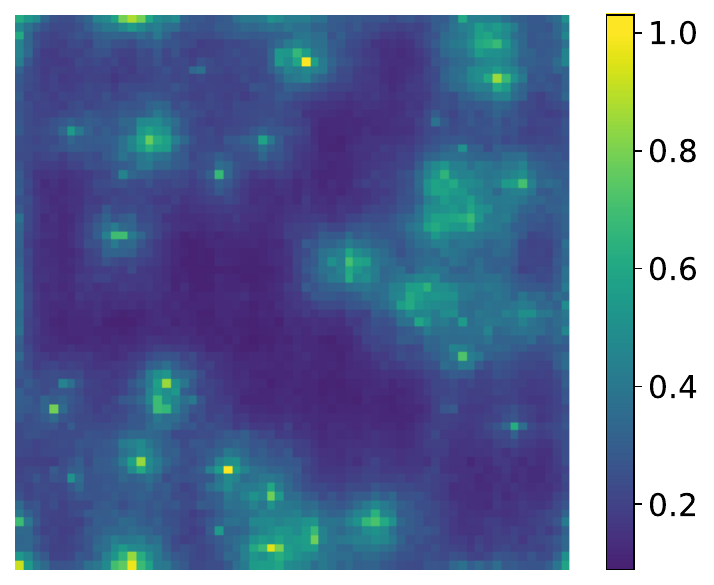}
        \vspace{-1.5\baselineskip}
        \caption{}
        \label{arb}
    \end{subfigure}%
    \hspace{10pt}%
    \begin{subfigure}[t]{0.21\textwidth}
        \includegraphics[width=1.0\textwidth]{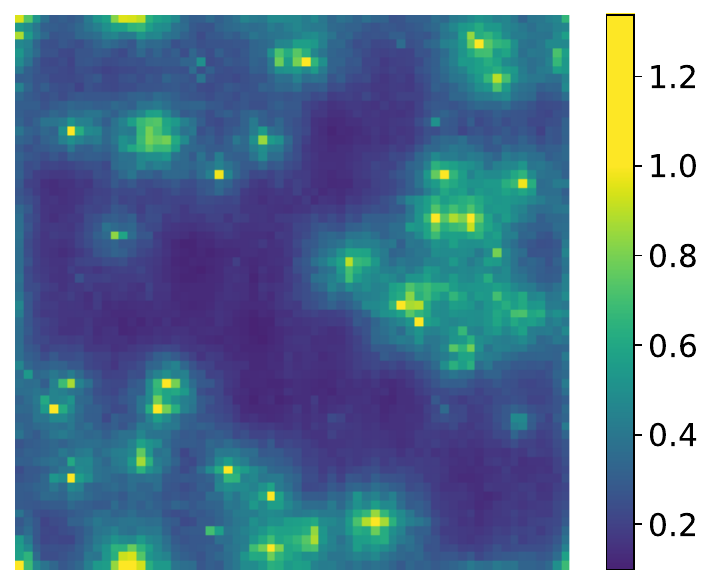}
        \vspace{-1.5\baselineskip}
        \caption{}
        \label{carb}
    \end{subfigure}
    
    \caption{Pixel-wise uncertainty quantification represented by the $90$-th quantile estimation. Subfigure (\subref{absres}) shows the oracle absolute residuals. Then, the estimation of each method is presented: (\subref{qr}) quantile regression, (\subref{cqr}) conformalized quantile regression, (\subref{pb}) parametric bootstrap, (\subref{eb}) equivariant bootstrap, (\subref{arb}) augmented radio bootstrap and (\subref{carb}) conformalized augmented radio bootstrap.}
    \label{fig:error}
    \vspace{-0.6\baselineskip}
\end{figure}

\begin{wraptable}[15]{l}{8.5cm}
\centering
\vspace{-1.0\intextsep}
    \caption{
    Uncertainty quantification performance comparison based on the average $\ell_2$ norm ratio between the $90\%$ confidence interval length and the ground truth image, and the empirical coverage percentage of the interval.
    }
    \vspace{-0.5\baselineskip}
    \label{tab:comp}
    \begin{tabular}{lcc}
        \toprule
        Method & Length ratio & Coverage \\
        \midrule
        Quantile Regression (QR) & 0.15 & 14\% \\
        Conformalized QR & 204.08 & 92\% \\
        Parametric Bootstrap & 0.07 & 0\% \\
        Equivariant Bootstrap & 0.13 & 7\% \\
        Augmented Radio Bootstrap & 0.29 & 87\% \\
        CARB & 0.34 & 91\% \\
        \bottomrule
    \end{tabular}
\end{wraptable}

The average time to construct confidence regions for each image is around 735ms for all bootstrapping techniques, compared to 153ms for the quantile regressions. The quantile regressions provide reduced times as it is a direct inference of the quantile. The draw of each bootstrap sample is independent of each other once the reconstruction $\hat{x}(y)$ is obtained. This fact allows us to exploit extremely easy parallelization techniques for the bootstrap sample generation, which we currently do by drawing $128$ samples in parallel. 

\begin{figure}
    \centering
    \includegraphics[width=0.5\linewidth]{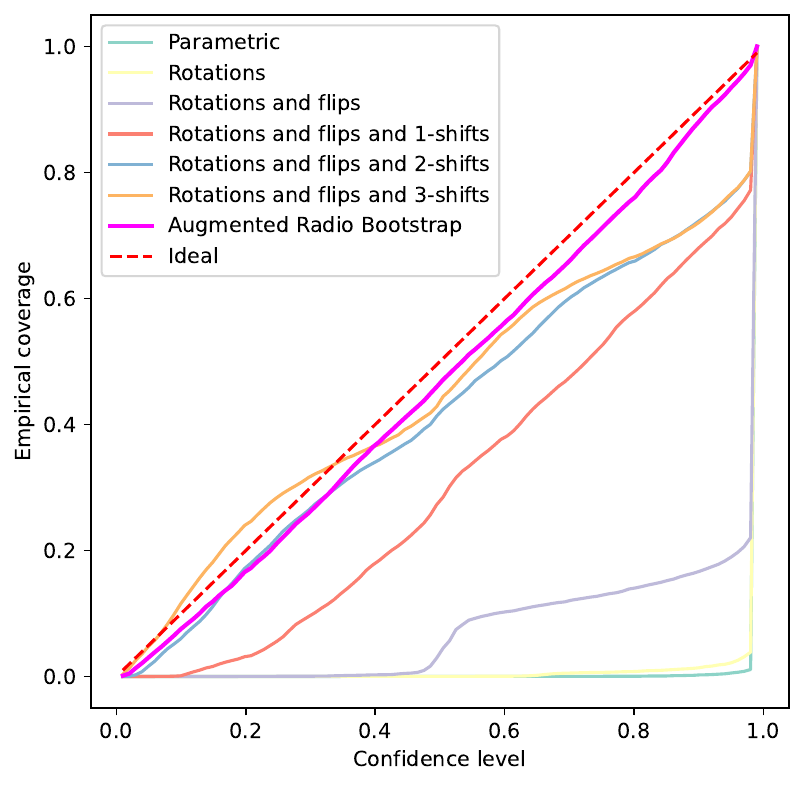}
    \caption{Coverage plots for the equivariant bootstrap methods with different group actions.}
    \label{fig:comparison}
\end{figure} 

A quantitative comparison of the UQ methods is presented in Table~\ref{tab:comp}. We first compare the average $\ell_2$ norm ratio between the confidence interval lengths and the ground truth image to study how tight the error estimations are. 
We then compare an empirical coverage of the confidence interval to verify if they are statistically valid. In other words, we empirically compute the expectation $\mathbb{E}_{y}\{ \text{P}(\hat{I}_{ \alpha}^{-}(y) < x^{\star} - \hat{x}(y)  < \hat{I}_{\alpha}^{+}(y)) | y \}$ over all pixels in the image dataset, where $\alpha$ is set to $0.1$, and $\hat{I}_{\alpha}^{-}(y), \hat{I}_{\alpha}^{+}(y)$ are the estimated lower and higher interval limits. The result should be greater or equal to $1-\delta$, with $\delta$ being the error rate. The parametric bootstrap significantly underestimates the errors, which is also true for the standard equivariant bootstrap. The error estimation greatly improves when adding the radio-spe cific transformations, showcasing the importance of setting problem-specific group actions. The subsequent conformalization procedure calibrates the estimated intervals such that the coverage verifies the required $1- \delta$ probability, allowing CARB to obtain tight bounds with correct coverage. We confirm that the CQR verifies the coverage probability but at the expense of too large uncertainty intervals, making the final UQ less informative.

Lastly, we present in Figure~\ref{fig:comparison} the coverage plots following \cite{Tachella2023} for the bootstrap-based methods and show how the curve changes as we include different group actions. We estimate a confidence region for $x^{\star}$, derived from the pivotal statistic $\lVert x^{\star} - \hat{x}(Y)\rVert_{2}^{2}$ related to the estimation's mean squared error. We then compute the empirical coverage probabilities on the test set, as measured by the proportion of test images that lie within the confidence regions for a range of confidence levels. The Augmented Radio Bootstrap method obtains an excellent coverage plot, although the probability is slightly underestimated. This result is remarkable in the context of the recent study \cite{Thong2024} showing that Bayesian imaging methods are not reporting reliable probabilities for UQ. 

One limitation of the current UQ method can be observed when the reconstruction algorithm used catastrophically fails to recover a feature from the ground truth image. In that situation, even if we draw bootstrap samples, the lost feature will not be recovered, and the uncertainty will be misestimated. Another limitation is the need for a representative calibration dataset for the conformalization procedure, which might not always be available in practice. In such a situation, the non-conformalized version of CARB should be used.

\begin{ack}

This work was performed using HPC resources from GENCI-IDRIS (Grant 2024-AD011015754). The code for CARB, as well as the trained models and the data used in this article, will be made publicly available in a future and more extensive publication.

Software used: \texttt{numpy} \cite{numpy}, \texttt{Pytorch} \cite{pytorch}, \texttt{Matplotlib} \cite{matplotlib}, \texttt{Jupyter} \cite{jupyter}, \texttt{Astropy} \cite{astropy}, \texttt{nenupy} \cite{nenupy}.
\end{ack}


\newpage
\clearpage
\small

\bibliography{bibliographie}

\begin{thebibliography}{59}
\providecommand{\natexlab}[1]{#1}
\providecommand{\url}[1]{\texttt{#1}}
\expandafter\ifx\csname urlstyle\endcsname\relax
  \providecommand{\doi}[1]{doi: #1}\else
  \providecommand{\doi}{doi: \begingroup \urlstyle{rm}\Url}\fi

\bibitem[Thompson et~al.(1991)Thompson, Moran, and Swenson]{Thompson2017}
Anthony Thompson, James Moran, and George Swenson, Jr.
\newblock \emph{Interferometry and Synthesis in Radio Astronomy}.
\newblock Astronomy and Astrophysics Library. Springer, 01 1991.
\newblock ISBN 978-3-319-44429-1.
\newblock \doi{10.1007/978-3-319-44431-4}.

\bibitem[Dewdney et~al.(2006)Dewdney, Hall, n~Schilizzi, and Lazio]{Dewdney2006}
Peter E.~F. Dewdney, Peter~J. Hall, Richard~T. n~Schilizzi, and T.J.L.W. Lazio.
\newblock The square kilometre array.
\newblock \emph{Proceedings of the IEEE}, 97:\penalty0 1482--1496, 2006.

\bibitem[Scaife(2020)]{scaife2020}
A.~M.~M. Scaife.
\newblock Big telescope, big data: towards exascale with the square kilometre array.
\newblock \emph{Philosophical Transactions of the Royal Society A: Mathematical, Physical and Engineering Sciences}, 378\penalty0 (2166):\penalty0 20190060, 2020.
\newblock \doi{10.1098/rsta.2019.0060}.
\newblock URL \url{https://royalsocietypublishing.org/doi/abs/10.1098/rsta.2019.0060}.

\bibitem[Högbom and Cornwell(1974)]{Hogbom1974}
J.~A. Högbom and Tim~J. Cornwell.
\newblock {Aperture Synthesis with a Non-Regular Distribution of Interferometer Baselines}.
\newblock \emph{Astronomy and Astrophysics}, 15:\penalty0 417--426, 1974.

\bibitem[{Offringa} et~al.(2014){Offringa}, {McKinley}, {Hurley-Walker}, {Briggs}, {Wayth}, {Kaplan}, {Bell}, {Feng}, {Neben}, {Hughes}, {Rhee}, {Murphy}, {Bhat}, {Bernardi}, {Bowman}, {Cappallo}, {Corey}, {Deshpande}, {Emrich}, {Ewall-Wice}, {Gaensler}, {Goeke}, {Greenhill}, {Hazelton}, {Hindson}, {Johnston-Hollitt}, {Jacobs}, {Kasper}, {Kratzenberg}, {Lenc}, {Lonsdale}, {Lynch}, {McWhirter}, {Mitchell}, {Morales}, {Morgan}, {Kudryavtseva}, {Oberoi}, {Ord}, {Pindor}, {Procopio}, {Prabu}, {Riding}, {Roshi}, {Shankar}, {Srivani}, {Subrahmanyan}, {Tingay}, {Waterson}, {Webster}, {Whitney}, {Williams}, and {Williams}]{offringa2014}
A.~R. {Offringa}, B.~{McKinley}, N.~{Hurley-Walker}, F.~H. {Briggs}, R.~B. {Wayth}, D.~L. {Kaplan}, M.~E. {Bell}, L.~{Feng}, A.~R. {Neben}, J.~D. {Hughes}, J.~{Rhee}, T.~{Murphy}, N.~D.~R. {Bhat}, G.~{Bernardi}, J.~D. {Bowman}, R.~J. {Cappallo}, B.~E. {Corey}, A.~A. {Deshpande}, D.~{Emrich}, A.~{Ewall-Wice}, B.~M. {Gaensler}, R.~{Goeke}, L.~J. {Greenhill}, B.~J. {Hazelton}, L.~{Hindson}, M.~{Johnston-Hollitt}, D.~C. {Jacobs}, J.~C. {Kasper}, E.~{Kratzenberg}, E.~{Lenc}, C.~J. {Lonsdale}, M.~J. {Lynch}, S.~R. {McWhirter}, D.~A. {Mitchell}, M.~F. {Morales}, E.~{Morgan}, N.~{Kudryavtseva}, D.~{Oberoi}, S.~M. {Ord}, B.~{Pindor}, P.~{Procopio}, T.~{Prabu}, J.~{Riding}, D.~A. {Roshi}, N.~Udaya {Shankar}, K.~S. {Srivani}, R.~{Subrahmanyan}, S.~J. {Tingay}, M.~{Waterson}, R.~L. {Webster}, A.~R. {Whitney}, A.~{Williams}, and C.~L. {Williams}.
\newblock {WSCLEAN: an implementation of a fast, generic wide-field imager for radio astronomy}.
\newblock \emph{Monthly Notices of the Royal Astronomical Society}, 444\penalty0 (1):\penalty0 606--619, October 2014.
\newblock \doi{10.1093/mnras/stu1368}.

\bibitem[{Offringa} and {Smirnov}(2017)]{offringa2017}
A.~R. {Offringa} and O.~{Smirnov}.
\newblock {An optimized algorithm for multiscale wideband deconvolution of radio astronomical images}.
\newblock \emph{Monthly Notices of the Royal Astronomical Society}, 471\penalty0 (1):\penalty0 301--316, October 2017.
\newblock \doi{10.1093/mnras/stx1547}.

\bibitem[McEwen and Wiaux(2011)]{mcewen2011}
J.~D. McEwen and Y.~Wiaux.
\newblock {Compressed sensing for wide-field radio interferometric imaging}.
\newblock \emph{Monthly Notices of the Royal Astronomical Society}, 413\penalty0 (2):\penalty0 1318--1332, 04 2011.
\newblock ISSN 0035-8711.
\newblock \doi{10.1111/j.1365-2966.2011.18217.x}.
\newblock URL \url{https://doi.org/10.1111/j.1365-2966.2011.18217.x}.

\bibitem[Carrillo et~al.(2012)Carrillo, McEwen, and Wiaux]{carrillo2012}
R.~E. Carrillo, J.~D. McEwen, and Y.~Wiaux.
\newblock {Sparsity Averaging Reweighted Analysis (SARA): a novel algorithm for radio-interferometric imaging}.
\newblock \emph{Monthly Notices of the Royal Astronomical Society}, 426\penalty0 (2):\penalty0 1223--1234, 10 2012.
\newblock ISSN 0035-8711.
\newblock \doi{10.1111/j.1365-2966.2012.21605.x}.
\newblock URL \url{https://doi.org/10.1111/j.1365-2966.2012.21605.x}.

\bibitem[Carrillo et~al.(2014)Carrillo, McEwen, and Wiaux]{carrillo2014}
R.~E. Carrillo, J.~D. McEwen, and Y.~Wiaux.
\newblock {PURIFY: a new approach to radio-interferometric imaging}.
\newblock \emph{Monthly Notices of the Royal Astronomical Society}, 439\penalty0 (4):\penalty0 3591--3604, 02 2014.
\newblock ISSN 0035-8711.
\newblock \doi{10.1093/mnras/stu202}.
\newblock URL \url{https://doi.org/10.1093/mnras/stu202}.

\bibitem[Dabbech et~al.(2018)Dabbech, Onose, Abdulaziz, Perley, Smirnov, and Wiaux]{dabbech2018}
A~Dabbech, A~Onose, A~Abdulaziz, R~A Perley, O~M Smirnov, and Y~Wiaux.
\newblock {Cygnus A super-resolved via convex optimization from VLA data}.
\newblock \emph{Monthly Notices of the Royal Astronomical Society}, 476\penalty0 (3):\penalty0 2853--2866, 02 2018.
\newblock ISSN 0035-8711.
\newblock \doi{10.1093/mnras/sty372}.
\newblock URL \url{https://doi.org/10.1093/mnras/sty372}.

\bibitem[{Junklewitz, H.} et~al.(2016){Junklewitz, H.}, {Bell, M. R.}, {Selig, M.}, and {En{\ss}lin, T. A.}]{junklewitz2016}
{Junklewitz, H.}, {Bell, M. R.}, {Selig, M.}, and {En{\ss}lin, T. A.}
\newblock Resolve: A new algorithm for aperture synthesis imaging of extended emission in radio astronomy.
\newblock \emph{{A\&A}}, 586:\penalty0 A76, 2016.
\newblock \doi{10.1051/0004-6361/201323094}.
\newblock URL \url{https://doi.org/10.1051/0004-6361/201323094}.

\bibitem[{Arras} et~al.(2018){Arras}, {Knollm{\"u}ller}, {Junklewitz}, and {En{\ss}lin}]{Arras2018}
Philipp {Arras}, Jakob {Knollm{\"u}ller}, Henrik {Junklewitz}, and Torsten~A. {En{\ss}lin}.
\newblock {Radio Imaging With Information Field Theory}.
\newblock \emph{arXiv e-prints}, art. arXiv:1803.02174, March 2018.
\newblock \doi{10.48550/arXiv.1803.02174}.

\bibitem[{Knollm{\"u}ller} and {En{\ss}lin}(2019)]{knollmuller2019}
Jakob {Knollm{\"u}ller} and Torsten~A. {En{\ss}lin}.
\newblock {Metric Gaussian Variational Inference}.
\newblock \emph{arXiv e-prints}, art. arXiv:1901.11033, January 2019.
\newblock \doi{10.48550/arXiv.1901.11033}.

\bibitem[{Arras} et~al.(2021){Arras}, {Bester}, {Perley, Richard A.}, {Leike, Reimar}, {Smirnov, Oleg}, {Westermann, R{\"u}diger}, and {En{\ss}lin, Torsten A.}]{arras2021}
Philipp {Arras}, Hertzog~L. {Bester}, {Perley, Richard A.}, {Leike, Reimar}, {Smirnov, Oleg}, {Westermann, R{\"u}diger}, and {En{\ss}lin, Torsten A.}
\newblock Comparison of classical and bayesian imaging in radio interferometry - cygnus a with clean and resolve.
\newblock \emph{{A\&A}}, 646:\penalty0 A84, 2021.
\newblock \doi{10.1051/0004-6361/202039258}.
\newblock URL \url{https://doi.org/10.1051/0004-6361/202039258}.

\bibitem[Terris et~al.(2022)Terris, Dabbech, Tang, and Wiaux]{terris2022}
Matthieu Terris, Arwa Dabbech, Chao Tang, and Yves Wiaux.
\newblock {Image reconstruction algorithms in radio interferometry: From handcrafted to learned regularization denoisers}.
\newblock \emph{Monthly Notices of the Royal Astronomical Society}, 518\penalty0 (1):\penalty0 604--622, 09 2022.
\newblock ISSN 0035-8711.
\newblock \doi{10.1093/mnras/stac2672}.
\newblock URL \url{https://doi.org/10.1093/mnras/stac2672}.

\bibitem[Aghabiglou et~al.(2024)Aghabiglou, Chu, Dabbech, and Wiaux]{Aghabiglou2024}
Amir Aghabiglou, Chung~San Chu, Arwa Dabbech, and Yves Wiaux.
\newblock {The R2D2 Deep Neural Network Series Paradigm for Fast Precision Imaging in Radio Astronomy}.
\newblock \emph{The Astrophysical Journal Supplement Series}, 273\penalty0 (1):\penalty0 3, 2024.
\newblock \doi{10.3847/1538-4365/ad46f5}.

\bibitem[{Mars} et~al.(2024){Mars}, {Betcke}, and {McEwen}]{mars2024}
Matthijs {Mars}, Marta~M. {Betcke}, and Jason~D. {McEwen}.
\newblock {Learned radio interferometric imaging for varying visibility coverage}.
\newblock \emph{arXiv e-prints}, art. arXiv:2405.08958, May 2024.
\newblock \doi{10.48550/arXiv.2405.08958}.

\bibitem[Chen et~al.(2022)Chen, Chen, Chen, Heaton, Liu, Wang, and Yin]{chen2022}
Tianlong Chen, Xiaohan Chen, Wuyang Chen, Howard Heaton, Jialin Liu, Zhangyang Wang, and Wotao Yin.
\newblock Learning to optimize: A primer and a benchmark.
\newblock \emph{Journal of Machine Learning Research}, 23\penalty0 (189):\penalty0 1--59, 2022.
\newblock URL \url{http://jmlr.org/papers/v23/21-0308.html}.

\bibitem[Kern et~al.(2024)Kern, Bobin, and Kervazo]{Kern2024}
Jonathan Kern, Jérôme Bobin, and Christophe Kervazo.
\newblock {EVIL-Deconv: Efficient Variability-Informed Learned Deconvolution using Algorithm Unrolling}.
\newblock \emph{[Manuscript submitted for publication]}, 2024.

\bibitem[{Cai} et~al.(2018{\natexlab{a}}){Cai}, {Pereyra}, and {McEwen}]{Cai2017a}
Xiaohao {Cai}, Marcelo {Pereyra}, and Jason~D. {McEwen}.
\newblock {Uncertainty quantification for radio interferometric imaging - I. Proximal MCMC methods}.
\newblock \emph{MNRAS}, 480\penalty0 (3):\penalty0 4154--4169, November 2018{\natexlab{a}}.
\newblock \doi{10.1093/mnras/sty2004}.

\bibitem[Angelopoulos and Bates(2021)]{Angelopoulos2021}
Anastasios~Nikolas Angelopoulos and Stephen Bates.
\newblock {A Gentle Introduction to Conformal Prediction and Distribution-Free Uncertainty Quantification}.
\newblock \emph{arXiv e-prints}, art. arXiv:2107.07511, July 2021.
\newblock \doi{10.48550/arXiv.2107.07511}.

\bibitem[Pereyra(2017)]{pereyra2017}
Marcelo Pereyra.
\newblock Maximum-a-posteriori estimation with bayesian confidence regions.
\newblock \emph{SIAM Journal on Imaging Sciences}, 10\penalty0 (1):\penalty0 285--302, 2017.
\newblock \doi{10.1137/16M1071249}.
\newblock URL \url{https://doi.org/10.1137/16M1071249}.

\bibitem[{Cai} et~al.(2018{\natexlab{b}}){Cai}, {Pereyra}, and {McEwen}]{Cai2017b}
Xiaohao {Cai}, Marcelo {Pereyra}, and Jason~D. {McEwen}.
\newblock {Uncertainty quantification for radio interferometric imaging: II. MAP estimation}.
\newblock \emph{MNRAS}, 480\penalty0 (3):\penalty0 4170--4182, November 2018{\natexlab{b}}.
\newblock \doi{10.1093/mnras/sty2015}.

\bibitem[Liaudat et~al.(2024)Liaudat, Mars, Price, Pereyra, Betcke, and McEwen]{Liaudat2023}
Tobías~I Liaudat, Matthijs Mars, Matthew~A Price, Marcelo Pereyra, Marta~M Betcke, and Jason~D McEwen.
\newblock {Scalable Bayesian uncertainty quantification with data-driven priors for radio interferometric imaging}.
\newblock \emph{RAS Techniques and Instruments}, 3\penalty0 (1):\penalty0 505--534, 08 2024.
\newblock ISSN 2752-8200.
\newblock \doi{10.1093/rasti/rzae030}.
\newblock URL \url{https://doi.org/10.1093/rasti/rzae030}.

\bibitem[{Terris} et~al.(2023){Terris}, {Tang}, {Jackson}, and {Wiaux}]{Terris2023}
Matthieu {Terris}, Chao {Tang}, Adrian {Jackson}, and Yves {Wiaux}.
\newblock {The AIRI plug-and-play algorithm for image reconstruction in radio-interferometry: variations and robustness}.
\newblock \emph{arXiv e-prints}, art. arXiv:2312.07137, December 2023.
\newblock \doi{10.48550/arXiv.2312.07137}.

\bibitem[{Aghabiglou} et~al.(2024){Aghabiglou}, {San Chu}, {Dabbech}, and {Wiaux}]{Aghabiglou2024UQ}
Amir {Aghabiglou}, Chung {San Chu}, Arwa {Dabbech}, and Yves {Wiaux}.
\newblock {R2D2 image reconstruction with model uncertainty quantification in radio astronomy}.
\newblock \emph{arXiv e-prints}, art. arXiv:2403.18052, March 2024.
\newblock \doi{10.48550/arXiv.2403.18052}.

\bibitem[{Dia} et~al.(2023){Dia}, {Yantovski-Barth}, {Adam}, {Bowles}, {Lemos}, {Scaife}, {Hezaveh}, and {Perreault-Levasseur}]{dia2023}
Noe {Dia}, M.~J. {Yantovski-Barth}, Alexandre {Adam}, Micah {Bowles}, Pablo {Lemos}, Anna M.~M. {Scaife}, Yashar {Hezaveh}, and Laurence {Perreault-Levasseur}.
\newblock {Bayesian Imaging for Radio Interferometry with Score-Based Priors}.
\newblock \emph{arXiv e-prints}, art. arXiv:2311.18012, November 2023.
\newblock \doi{10.48550/arXiv.2311.18012}.

\bibitem[Sun et~al.(2024)Sun, Wu, Chen, Feng, and Bouman]{sun2024}
Yu~Sun, Zihui Wu, Yifan Chen, Berthy~T. Feng, and Katherine~L. Bouman.
\newblock Provable probabilistic imaging using score-based generative priors.
\newblock \emph{IEEE Transactions on Computational Imaging}, 10:\penalty0 1290--1305, 2024.
\newblock \doi{10.1109/TCI.2024.3449114}.

\bibitem[{Wu} et~al.(2024){Wu}, {Sun}, {Chen}, {Zhang}, {Yue}, and {Bouman}]{wu2024}
Zihui {Wu}, Yu~{Sun}, Yifan {Chen}, Bingliang {Zhang}, Yisong {Yue}, and Katherine~L. {Bouman}.
\newblock {Principled Probabilistic Imaging using Diffusion Models as Plug-and-Play Priors}.
\newblock \emph{arXiv e-prints}, art. arXiv:2405.18782, May 2024.
\newblock \doi{10.48550/arXiv.2405.18782}.

\bibitem[Venkatakrishnan et~al.(2013)Venkatakrishnan, Bouman, and Wohlberg]{venkatakrishnan2013}
Singanallur~V. Venkatakrishnan, Charles~A. Bouman, and Brendt Wohlberg.
\newblock Plug-and-play priors for model based reconstruction.
\newblock In \emph{2013 IEEE Global Conference on Signal and Information Processing}, pages 945--948, 2013.
\newblock \doi{10.1109/GlobalSIP.2013.6737048}.

\bibitem[Gregor and LeCun(2010)]{Gregor2010}
Karol Gregor and Yann LeCun.
\newblock Learning fast approximations of sparse coding.
\newblock In \emph{Proceedings of the 27th International Conference on International Conference on Machine Learning}, ICML'10, page 399–406, Madison, WI, USA, 2010. Omnipress.
\newblock ISBN 9781605589077.

\bibitem[Monga et~al.(2021)Monga, Li, and Eldar]{Monga2021}
Vishal Monga, Yuelong Li, and Yonina~C. Eldar.
\newblock Algorithm unrolling: Interpretable, efficient deep learning for signal and image processing.
\newblock \emph{IEEE Signal Processing Magazine}, 38\penalty0 (2):\penalty0 18--44, 2021.
\newblock \doi{10.1109/MSP.2020.3016905}.

\bibitem[Beck and Teboulle(2009)]{Beck2009}
Amir Beck and Marc Teboulle.
\newblock A fast iterative shrinkage-thresholding algorithm for linear inverse problems.
\newblock \emph{SIAM Journal on Imaging Sciences}, 2\penalty0 (1):\penalty0 183--202, 2009.
\newblock \doi{10.1137/080716542}.
\newblock URL \url{https://doi.org/10.1137/080716542}.

\bibitem[{Chen} et~al.(2018){Chen}, {Liu}, {Wang}, and {Yin}]{Chen2018}
Xiaohan {Chen}, Jialin {Liu}, Zhangyang {Wang}, and Wotao {Yin}.
\newblock {Theoretical Linear Convergence of Unfolded ISTA and its Practical Weights and Thresholds}.
\newblock \emph{Advances in Neural Information Processing Systems}, 31:\penalty0 arXiv:1808.10038, August 2018.
\newblock \doi{10.48550/arXiv.1808.10038}.

\bibitem[{Zhang} et~al.(2020){Zhang}, {Li}, {Zuo}, {Zhang}, {Van Gool}, and {Timofte}]{Zhang2020}
Kai {Zhang}, Yawei {Li}, Wangmeng {Zuo}, Lei {Zhang}, Luc {Van Gool}, and Radu {Timofte}.
\newblock {Plug-and-Play Image Restoration with Deep Denoiser Prior}.
\newblock \emph{IEEE Transactions on Pattern Analysis and Machine Intelligence}, August 2020.
\newblock \doi{10.48550/arXiv.2008.13751}.

\bibitem[Pereyra and Tachella(2024)]{Tachella2023}
Marcelo Pereyra and Juli\'{a}n Tachella.
\newblock Equivariant bootstrapping for uncertainty quantification in imaging inverse problems.
\newblock In \emph{Proceedings of The 27th International Conference on Artificial Intelligence and Statistics}, volume 238 of \emph{Proceedings of Machine Learning Research}, pages 4141--4149. PMLR, 02--04 May 2024.
\newblock URL \url{https://proceedings.mlr.press/v238/pereyra24a.html}.

\bibitem[Chen et~al.(2021)Chen, Tachella, and Davies]{chen2021}
Dongdong Chen, Julián Tachella, and Mike~E. Davies.
\newblock Equivariant imaging: Learning beyond the range space.
\newblock In \emph{2021 IEEE/CVF International Conference on Computer Vision (ICCV)}, pages 4359--4368, 2021.
\newblock \doi{10.1109/ICCV48922.2021.00434}.

\bibitem[Tachella et~al.(2024)Tachella, Chen, and Davies]{Tachella2022}
Juli\'{a}n Tachella, Dongdong Chen, and Mike Davies.
\newblock Sensing theorems for unsupervised learning in linear inverse problems.
\newblock \emph{J. Mach. Learn. Res.}, 24\penalty0 (1), mar 2024.
\newblock ISSN 1532-4435.
\newblock \doi{10.48550/arXiv.2203.12513}.

\bibitem[Angelopoulos et~al.(2022)Angelopoulos, Kohli, Bates, Jordan, Malik, Alshaabi, Upadhyayula, and Romano]{Angelopoulos2022}
Anastasios~N Angelopoulos, Amit~Pal Kohli, Stephen Bates, Michael Jordan, Jitendra Malik, Thayer Alshaabi, Srigokul Upadhyayula, and Yaniv Romano.
\newblock Image-to-image regression with distribution-free uncertainty quantification and applications in imaging.
\newblock In Kamalika Chaudhuri, Stefanie Jegelka, Le~Song, Csaba Szepesvari, Gang Niu, and Sivan Sabato, editors, \emph{Proceedings of the 39th International Conference on Machine Learning}, volume 162 of \emph{Proceedings of Machine Learning Research}, pages 717--730. PMLR, 17--23 Jul 2022.
\newblock URL \url{https://proceedings.mlr.press/v162/angelopoulos22a.html}.

\bibitem[Koenker and Bassett(1978)]{Koenker1978}
Roger Koenker and Gilbert Bassett.
\newblock Regression quantiles.
\newblock \emph{Econometrica}, 46\penalty0 (1):\penalty0 33--50, 1978.
\newblock ISSN 00129682, 14680262.
\newblock URL \url{http://www.jstor.org/stable/1913643}.

\bibitem[Romano et~al.(2019)Romano, Patterson, and Candes]{Romano2019}
Yaniv Romano, Evan Patterson, and Emmanuel Candes.
\newblock Conformalized quantile regression.
\newblock \emph{Advances in neural information processing systems}, 32, 2019.
\newblock \doi{10.48550/arXiv.1905.03222}.

\bibitem[Observatory(2016)]{Meerkat}
South African Radio~Astronomy Observatory.
\newblock {MeerKAT}, 2016.
\newblock URL \url{https://public.ska.ac.za/meerkat}.

\bibitem[{Thong} et~al.(2024){Thong}, {Kemajou Mbakam}, and {Pereyra}]{Thong2024}
David Y.~W. {Thong}, Charlesquin {Kemajou Mbakam}, and Marcelo {Pereyra}.
\newblock {Do Bayesian imaging methods report trustworthy probabilities?}
\newblock \emph{arXiv e-prints}, art. arXiv:2405.08179, May 2024.
\newblock \doi{10.48550/arXiv.2405.08179}.

\bibitem[Harris et~al.(2020)Harris, Millman, van~der Walt, Gommers, Virtanen, Cournapeau, Wieser, Taylor, Berg, Smith, Kern, Picus, Hoyer, van Kerkwijk, Brett, Haldane, del R{\'{i}}o, Wiebe, Peterson, G{\'{e}}rard-Marchant, Sheppard, Reddy, Weckesser, Abbasi, Gohlke, and Oliphant]{numpy}
Charles~R. Harris, K.~Jarrod Millman, St{\'{e}}fan~J. van~der Walt, Ralf Gommers, Pauli Virtanen, David Cournapeau, Eric Wieser, Julian Taylor, Sebastian Berg, Nathaniel~J. Smith, Robert Kern, Matti Picus, Stephan Hoyer, Marten~H. van Kerkwijk, Matthew Brett, Allan Haldane, Jaime~Fern{\'{a}}ndez del R{\'{i}}o, Mark Wiebe, Pearu Peterson, Pierre G{\'{e}}rard-Marchant, Kevin Sheppard, Tyler Reddy, Warren Weckesser, Hameer Abbasi, Christoph Gohlke, and Travis~E. Oliphant.
\newblock Array programming with {NumPy}.
\newblock \emph{Nature}, 585\penalty0 (7825):\penalty0 357--362, September 2020.
\newblock \doi{10.1038/s41586-020-2649-2}.
\newblock URL \url{https://doi.org/10.1038/s41586-020-2649-2}.

\bibitem[{Paszke} et~al.(2019){Paszke}, {Gross}, {Massa}, {Lerer}, {Bradbury}, {Chanan}, {Killeen}, {Lin}, {Gimelshein}, {Antiga}, {Desmaison}, {K{\"o}pf}, {Yang}, {DeVito}, {Raison}, {Tejani}, {Chilamkurthy}, {Steiner}, {Fang}, {Bai}, and {Chintala}]{pytorch}
Adam {Paszke}, Sam {Gross}, Francisco {Massa}, Adam {Lerer}, James {Bradbury}, Gregory {Chanan}, Trevor {Killeen}, Zeming {Lin}, Natalia {Gimelshein}, Luca {Antiga}, Alban {Desmaison}, Andreas {K{\"o}pf}, Edward {Yang}, Zach {DeVito}, Martin {Raison}, Alykhan {Tejani}, Sasank {Chilamkurthy}, Benoit {Steiner}, Lu~{Fang}, Junjie {Bai}, and Soumith {Chintala}.
\newblock {PyTorch: An Imperative Style, High-Performance Deep Learning Library}.
\newblock \emph{arXiv e-prints}, art. arXiv:1912.01703, December 2019.
\newblock \doi{10.48550/arXiv.1912.01703}.

\bibitem[Hunter(2007)]{matplotlib}
J.~D. Hunter.
\newblock Matplotlib: A 2d graphics environment.
\newblock \emph{Computing in Science \& Engineering}, 9\penalty0 (3):\penalty0 90--95, 2007.
\newblock \doi{10.1109/MCSE.2007.55}.

\bibitem[Kluyver et~al.(2016)Kluyver, Ragan-Kelley, P{\'e}rez, Granger, Bussonnier, Frederic, Kelley, Hamrick, Grout, Corlay, Ivanov, Avila, Abdalla, Willing, and development team]{jupyter}
Thomas Kluyver, Benjamin Ragan-Kelley, Fernando P{\'e}rez, Brian Granger, Matthias Bussonnier, Jonathan Frederic, Kyle Kelley, Jessica Hamrick, Jason Grout, Sylvain Corlay, Paul Ivanov, Dami{\'a}n Avila, Safia Abdalla, Carol Willing, and Jupyter development team.
\newblock Jupyter notebooks ? a publishing format for reproducible computational workflows.
\newblock In Fernando Loizides and Birgit Scmidt, editors, \emph{Positioning and Power in Academic Publishing: Players, Agents and Agendas}, pages 87--90. IOS Press, 2016.
\newblock URL \url{https://eprints.soton.ac.uk/403913/}.

\bibitem[{Astropy Collaboration} et~al.(2022){Astropy Collaboration}, {Price-Whelan}, {Lim}, {Earl}, {Starkman}, {Bradley}, {Shupe}, {Patil}, {Corrales}, {Brasseur}, {N{\"o}the}, {Donath}, {Tollerud}, {Morris}, {Ginsburg}, {Vaher}, {Weaver}, {Tocknell}, {Jamieson}, {van Kerkwijk}, {Robitaille}, {Merry}, {Bachetti}, {G{\"u}nther}, {Aldcroft}, {Alvarado-Montes}, {Archibald}, {B{\'o}di}, {Bapat}, {Barentsen}, {Baz{\'a}n}, {Biswas}, {Boquien}, {Burke}, {Cara}, {Cara}, {Conroy}, {Conseil}, {Craig}, {Cross}, {Cruz}, {D'Eugenio}, {Dencheva}, {Devillepoix}, {Dietrich}, {Eigenbrot}, {Erben}, {Ferreira}, {Foreman-Mackey}, {Fox}, {Freij}, {Garg}, {Geda}, {Glattly}, {Gondhalekar}, {Gordon}, {Grant}, {Greenfield}, {Groener}, {Guest}, {Gurovich}, {Handberg}, {Hart}, {Hatfield-Dodds}, {Homeier}, {Hosseinzadeh}, {Jenness}, {Jones}, {Joseph}, {Kalmbach}, {Karamehmetoglu}, {Ka{\l}uszy{\'n}ski}, {Kelley}, {Kern}, {Kerzendorf}, {Koch}, {Kulumani}, {Lee}, {Ly}, {Ma}, {MacBride}, {Maljaars}, {Muna}, {Murphy}, {Norman}, {O'Steen}, {Oman}, {Pacifici}, {Pascual}, {Pascual-Granado}, {Patil}, {Perren}, {Pickering}, {Rastogi}, {Roulston}, {Ryan}, {Rykoff}, {Sabater}, {Sakurikar}, {Salgado}, {Sanghi}, {Saunders}, {Savchenko}, {Schwardt}, {Seifert-Eckert}, {Shih}, {Jain}, {Shukla}, {Sick}, {Simpson}, {Singanamalla}, {Singer}, {Singhal}, {Sinha}, {Sip{\H{o}}cz}, {Spitler}, {Stansby}, {Streicher}, {{\v{S}}umak}, {Swinbank}, {Taranu}, {Tewary}, {Tremblay}, {de Val-Borro}, {Van Kooten}, {Vasovi{\'c}}, {Verma}, {de Miranda Cardoso}, {Williams}, {Wilson}, {Winkel}, {Wood-Vasey}, {Xue}, {Yoachim}, {Zhang}, {Zonca}, and {Astropy Project Contributors}]{astropy}
{Astropy Collaboration}, Adrian~M. {Price-Whelan}, Pey~Lian {Lim}, Nicholas {Earl}, Nathaniel {Starkman}, Larry {Bradley}, David~L. {Shupe}, Aarya~A. {Patil}, Lia {Corrales}, C.~E. {Brasseur}, Maximilian {N{\"o}the}, Axel {Donath}, Erik {Tollerud}, Brett~M. {Morris}, Adam {Ginsburg}, Eero {Vaher}, Benjamin~A. {Weaver}, James {Tocknell}, William {Jamieson}, Marten~H. {van Kerkwijk}, Thomas~P. {Robitaille}, Bruce {Merry}, Matteo {Bachetti}, H.~Moritz {G{\"u}nther}, Thomas~L. {Aldcroft}, Jaime~A. {Alvarado-Montes}, Anne~M. {Archibald}, Attila {B{\'o}di}, Shreyas {Bapat}, Geert {Barentsen}, Juanjo {Baz{\'a}n}, Manish {Biswas}, M{\'e}d{\'e}ric {Boquien}, D.~J. {Burke}, Daria {Cara}, Mihai {Cara}, Kyle~E. {Conroy}, Simon {Conseil}, Matthew~W. {Craig}, Robert~M. {Cross}, Kelle~L. {Cruz}, Francesco {D'Eugenio}, Nadia {Dencheva}, Hadrien A.~R. {Devillepoix}, J{\"o}rg~P. {Dietrich}, Arthur~Davis {Eigenbrot}, Thomas {Erben}, Leonardo {Ferreira}, Daniel {Foreman-Mackey}, Ryan {Fox}, Nabil {Freij}, Suyog {Garg}, Robel {Geda}, Lauren {Glattly}, Yash {Gondhalekar}, Karl~D. {Gordon}, David {Grant}, Perry {Greenfield}, Austen~M. {Groener}, Steve {Guest}, Sebastian {Gurovich}, Rasmus {Handberg}, Akeem {Hart}, Zac {Hatfield-Dodds}, Derek {Homeier}, Griffin {Hosseinzadeh}, Tim {Jenness}, Craig~K. {Jones}, Prajwel {Joseph}, J.~Bryce {Kalmbach}, Emir {Karamehmetoglu}, Miko{\l}aj {Ka{\l}uszy{\'n}ski}, Michael S.~P. {Kelley}, Nicholas {Kern}, Wolfgang~E. {Kerzendorf}, Eric~W. {Koch}, Shankar {Kulumani}, Antony {Lee}, Chun {Ly}, Zhiyuan {Ma}, Conor {MacBride}, Jakob~M. {Maljaars}, Demitri {Muna}, N.~A. {Murphy}, Henrik {Norman}, Richard {O'Steen}, Kyle~A. {Oman}, Camilla {Pacifici}, Sergio {Pascual}, J.~{Pascual-Granado}, Rohit~R. {Patil}, Gabriel~I. {Perren}, Timothy~E. {Pickering}, Tanuj {Rastogi}, Benjamin~R. {Roulston}, Daniel~F. {Ryan}, Eli~S. {Rykoff}, Jose {Sabater}, Parikshit {Sakurikar}, Jes{\'u}s {Salgado}, Aniket {Sanghi}, Nicholas {Saunders}, Volodymyr {Savchenko}, Ludwig {Schwardt}, Michael {Seifert-Eckert}, Albert~Y. {Shih}, Anany~Shrey {Jain}, Gyanendra {Shukla}, Jonathan {Sick}, Chris {Simpson}, Sudheesh {Singanamalla}, Leo~P. {Singer}, Jaladh {Singhal}, Manodeep {Sinha}, Brigitta~M. {Sip{\H{o}}cz}, Lee~R. {Spitler}, David {Stansby}, Ole {Streicher}, Jani {{\v{S}}umak}, John~D. {Swinbank}, Dan~S. {Taranu}, Nikita {Tewary}, Grant~R. {Tremblay}, Miguel {de Val-Borro}, Samuel~J. {Van Kooten}, Zlatan {Vasovi{\'c}}, Shresth {Verma}, Jos{\'e}~Vin{\'\i}cius {de Miranda Cardoso}, Peter K.~G. {Williams}, Tom~J. {Wilson}, Benjamin {Winkel}, W.~M. {Wood-Vasey}, Rui {Xue}, Peter {Yoachim}, Chen {Zhang}, Andrea {Zonca}, and {Astropy Project Contributors}.
\newblock {The Astropy Project: Sustaining and Growing a Community-oriented Open-source Project and the Latest Major Release (v5.0) of the Core Package}.
\newblock \emph{ApJ}, 935\penalty0 (2):\penalty0 167, August 2022.
\newblock \doi{10.3847/1538-4357/ac7c74}.

\bibitem[{Alan Loh, Julien Girard, {and} the NenuFAR team}(2020)]{nenupy}
{Alan Loh, Julien Girard, {and} the NenuFAR team}.
\newblock nenupy: a python package for the low-frequency radio telescope nenufar, November 2020.
\newblock URL \url{https://doi.org/10.5281/zenodo.3667815}.

\bibitem[Tibshirani(1996)]{Tibshirani1996}
Robert Tibshirani.
\newblock Regression shrinkage and selection via the lasso.
\newblock \emph{Journal of the Royal Statistical Society. Series B (Methodological)}, 58\penalty0 (1):\penalty0 267--288, 1996.
\newblock ISSN 00359246.
\newblock URL \url{http://www.jstor.org/stable/2346178}.

\bibitem[Ablin et~al.(2019)Ablin, Moreau, Massias, and Gramfort]{Ablin2019}
Pierre Ablin, Thomas Moreau, Mathurin Massias, and Alexandre Gramfort.
\newblock {Learning step sizes for unfolded sparse coding}.
\newblock \emph{Neural Information Processing Systems}, 32, 2019.
\newblock \doi{10.48550/arXiv.1905.11071}.

\bibitem[Tichonov et~al.(1977)Tichonov, John, and Arsenin]{Tikhonov1977}
A.N. Tichonov, F.~John, and V.J. Arsenin.
\newblock \emph{Solutions of Ill-posed Problems}.
\newblock Halsted Press book. V.H. Winston \& Sons, 1977.

\bibitem[Ma et~al.(2022)Ma, Yan, Li, and Zhao]{Ma2022}
Ge~Ma, Ziwei Yan, Zhifu Li, and Zhijia Zhao.
\newblock Efficient iterative regularization method for total variation-based image restoration.
\newblock \emph{Electronics}, 11\penalty0 (2), 2022.
\newblock ISSN 2079-9292.
\newblock \doi{10.3390/electronics11020258}.

\bibitem[{Zhang} et~al.(2017){Zhang}, {Zuo}, {Gu}, and {Zhang}]{Zhang2017}
Kai {Zhang}, Wangmeng {Zuo}, Shuhang {Gu}, and Lei {Zhang}.
\newblock {Learning Deep CNN Denoiser Prior for Image Restoration}.
\newblock \emph{Proceedings of the IEEE conference on computer vision and pattern recognition}, pages 3929--3938, 2017.
\newblock \doi{10.48550/arXiv.1704.03264}.

\bibitem[Parikh and Boyd(2014)]{parikh2014}
Neal Parikh and Stephen Boyd.
\newblock Proximal algorithms.
\newblock \emph{Foundations and Trends® in Optimization}, 1\penalty0 (3):\penalty0 127--239, 2014.
\newblock ISSN 2167-3888.
\newblock \doi{10.1561/2400000003}.
\newblock URL \url{http://dx.doi.org/10.1561/2400000003}.

\bibitem[Zhang et~al.(2017)Zhang, Zuo, Chen, Meng, and Zhang]{Zhang2016}
Kai Zhang, Wangmeng Zuo, Yunjin Chen, Deyu Meng, and Lei Zhang.
\newblock Beyond a gaussian denoiser: Residual learning of deep cnn for image denoising.
\newblock \emph{Trans. Img. Proc.}, 26\penalty0 (7):\penalty0 3142–3155, jul 2017.
\newblock ISSN 1057-7149.
\newblock \doi{10.1109/TIP.2017.2662206}.
\newblock URL \url{https://doi.org/10.1109/TIP.2017.2662206}.

\bibitem[Pesquet et~al.(2020)Pesquet, Repetti, Terris, and Wiaux]{Pesquet2020}
Jean-Christophe Pesquet, Audrey Repetti, Matthieu Terris, and Yves Wiaux.
\newblock {Learning Maximally Monotone Operators for Image Recovery}.
\newblock \emph{SIAM J. Imaging Sci.}, 14:\penalty0 1206--1237, 2020.
\newblock \doi{10.48550/arXiv.2012.13247}.

\bibitem[Nix and Weigend(1994)]{Nix1994}
D.A. Nix and A.S. Weigend.
\newblock Estimating the mean and variance of the target probability distribution.
\newblock In \emph{Proceedings of 1994 IEEE International Conference on Neural Networks (ICNN'94)}, volume~1, pages 55--60 vol.1, 1994.
\newblock \doi{10.1109/ICNN.1994.374138}.

\bibitem[Rottmann et~al.(2019)Rottmann, Colling, Hack, Chan, Hüger, Schlicht, and Gottschalk]{Rottmann2019}
Matthias Rottmann, Pascal Colling, Thomas-Paul Hack, Robin Chan, Fabian Hüger, Peter Schlicht, and Hanno Gottschalk.
\newblock Prediction error meta classification in semantic segmentation: Detection via aggregated dispersion measures of softmax probabilities, 2019.
\newblock URL \url{https://arxiv.org/abs/1811.00648}.

\end{thebibliography}


\newpage

\appendix

\section{Radio-interferometry image reconstruction using an unrolled algorithm}
\subsection{Unrolling method}
\label{Unrolling}

Finding the image $x^{\star} \in \mathcal{X}$ that minimizes the reconstruction error with the observed radio-interferometry data is an ill-posed problem due to the form of the convolution operator $M$, or PSF, and the observational noise. Consequently, it is expected to add a regularization term to the optimization problem, which leads to solving a problem of the following form,
\begin{equation}
    \label{eq_reg}
    \Tilde{x} = \argmin_{x \in \mathcal{X}} \frac{1}{2}  \Vert y- M \ast x \Vert^2_2 + \lambda \mathcal{R}(x) ,
\end{equation}
where the regularization $\mathcal{R}$, can take the form of a simple sparsity constraint using a $l_1$ norm~\cite{Tibshirani1996}, the total variation function\cite{Ablin2019, Tikhonov1977, Ma2022} or a more complex function by relying on neural-network-based learned denoisers~\cite{Zhang2017, terris2022}.

Several iterative methods have been designed to solve this optimization problem, such as proximal gradient descent algorithms~\cite{Beck2009}, where each iteration can be computed as
\begin{equation}
\label{eq:PGD}
    x^{l+1}= \underbrace{\text{prox}_{\lambda \alpha}}_{\substack{\text{regularization}}} (\underbrace{x^l - \alpha M^*(y-Mx^l)}_{\substack{\text{data fidelity gradient}}}),
\end{equation}
where prox$()$ is the proximal operator \cite{parikh2014} associated with the chosen regularization function and $\alpha$ is the step size. In this work we follow the EVIL-Deconv method~\cite{Kern2024}, and we unroll $10$ iterations of Equation~\eqref{eq:PGD}  resulting in a neural network architecture with $10$ layers. Each layer can be expressed as
\begin{equation}
    x^{l+1}= p_l(x^l+\Phi_l(M)(y-M\ast x^l)),
\end{equation}
where $p_l$ and $\Phi_l$ are trainable operators. In practice, $p_l$ is a pretrained DRUNET \cite{Zhang2020} and $\Phi_l$ is a convolutional neural network.

\subsection{Unrolling results}
\label{app:table}

In this section, we present performance results obtained using the unrolling architecture described in the previous subsection, EVIL-Deconv \cite{Kern2024}, and compare it with two iterative methods, the CLEAN algorithm~\cite{Hogbom1974} and a Plug-and-Play (PnP) method, where the denoiser, a DnCNN \cite{Zhang2016}, is trained with a non-expansiveness constraint following~\cite{terris2022} to ensure convergence \cite{Pesquet2020}. We study two reconstruction performance metrics: the Normalized Mean Squared Error (NMSE), which we define as follows
\begin{equation}
    \text{NMSE [-dB]} = - 20 \log_{10} \left( \frac{\|x^{\star} - \hat{x}\|_2}{\| x^{\star} \|_2} \right) ,
\end{equation}
and the Structural Similarity Index Measure (SSIM). The mean reconstruction time for each method is also included to show the computational advantage of algorithm unrolling over iterative methods. Table~\ref{table_Kern} presents the performance results, which were computed on the data described in Section~\ref{sec:data}. All numerical experiments were done on two Nvidia Tesla V100 SXM2 16Gb GPUs.

\begin{table}[h]
    \centering
    \caption{Performance comparison of RI image reconstruction algorithms.}
    \label{table_Kern}
    \begin{tabular}{lccc}
        \toprule
        Method & Median NMSE [-dB] $\uparrow$  & SSIM $\uparrow$ & Time [ms] $\downarrow$ \\
        \midrule
        CLEAN & $4.2$ & $0.296$ & $794$ \\
        PnP  & $16.3$ & $0.869$ & $1110$ \\
        EVIL-Deconv & $\mathbf{19.9}$ & $\mathbf{0.970}$ & $\mathbf{51}$ \\
        \bottomrule
    \end{tabular}
\end{table}

\section{Point spread function and uv-coverage}
\label{psf_uv_coverage}

\begin{figure}[h]

    \centering
    \begin{subfigure}[t]{0.20\textwidth}
        \raggedleft
        \includegraphics[height=0.87\textwidth]{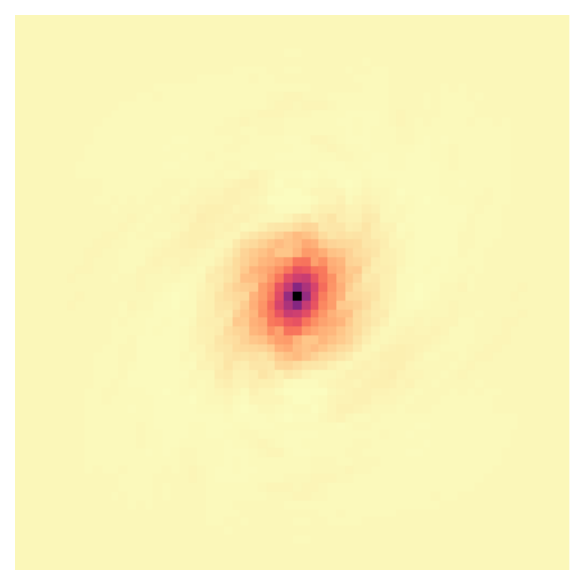}
    \end{subfigure}%
    \hspace{10pt}%
    \begin{subfigure}[t]{0.20\textwidth}
        \raggedleft
        \includegraphics[height=0.87\textwidth]{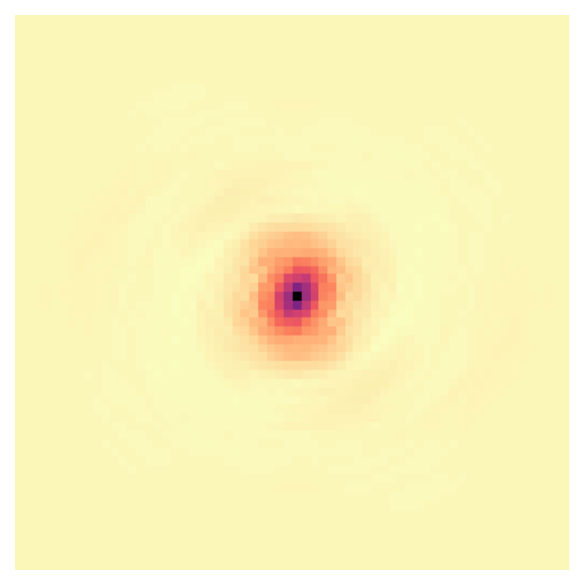}
    \end{subfigure}%
    \hspace{10pt}%
    \begin{subfigure}[t]{0.20\textwidth}
        \raggedleft
        \includegraphics[height=0.87\textwidth]{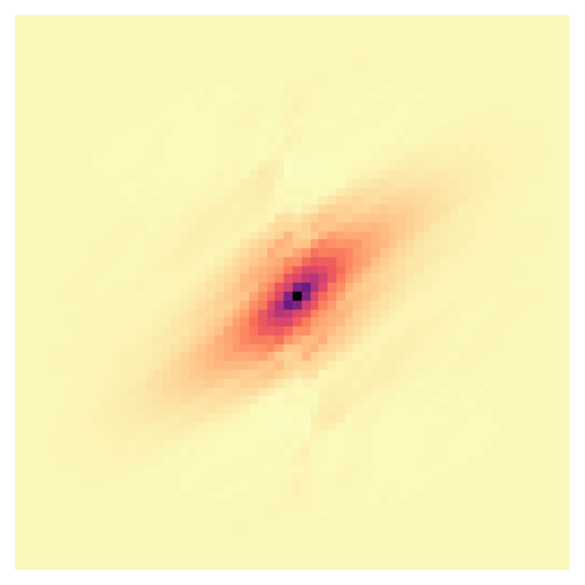}
    \end{subfigure}%
    \hspace{10pt}%
    \begin{subfigure}[t]{0.20\textwidth}
        \raggedleft
        \includegraphics[height=0.87\textwidth]{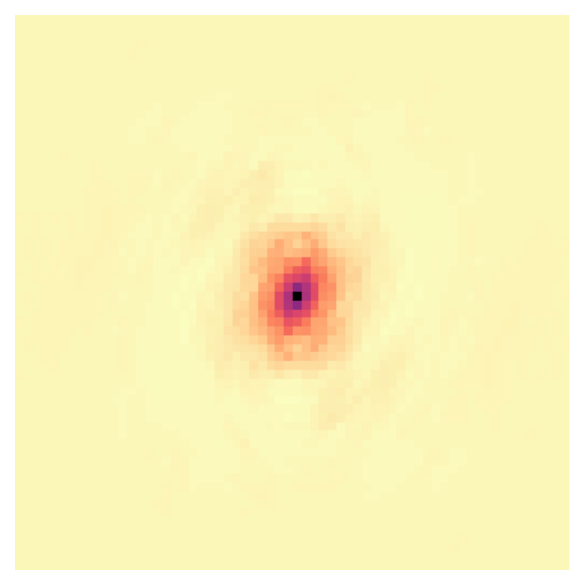}
    \end{subfigure}

    \centering
    \begin{subfigure}[t]{0.20\textwidth}
        \includegraphics[height=1\textwidth]{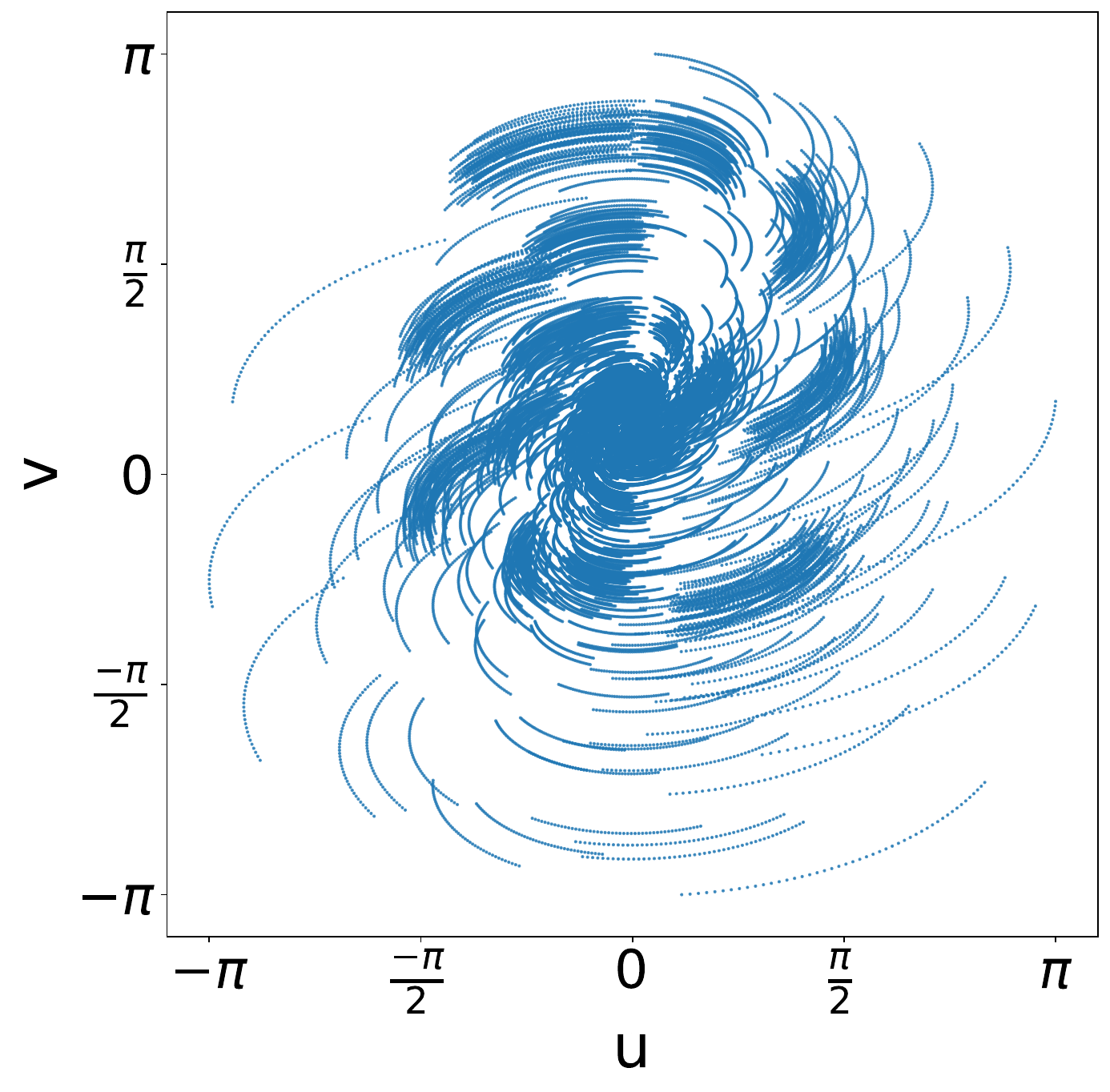}
    \end{subfigure}%
    \hspace{10pt}%
    \begin{subfigure}[t]{0.20\textwidth}
        \includegraphics[height=1\textwidth]{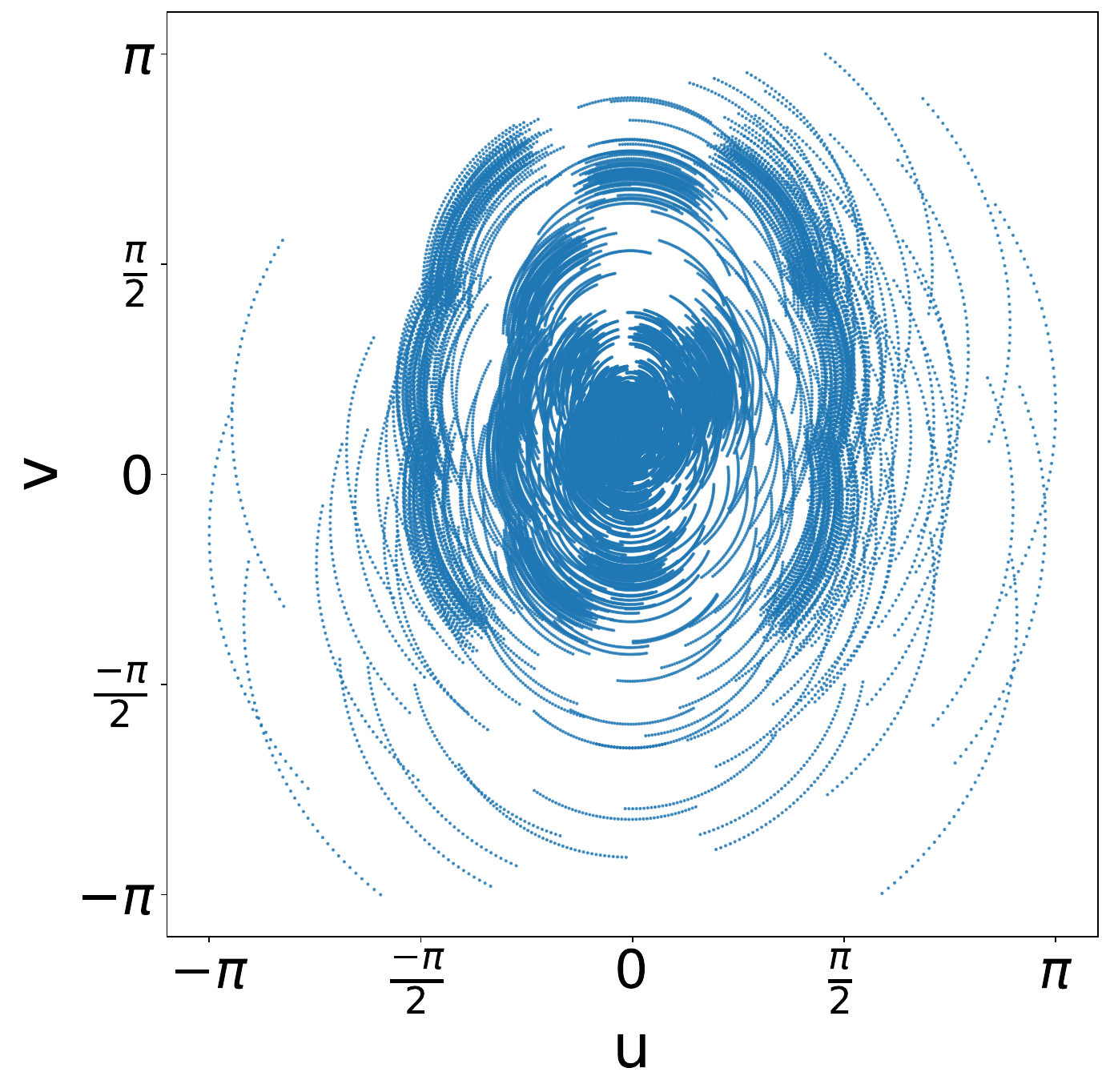}
    \end{subfigure}%
    \hspace{10pt}%
    \begin{subfigure}[t]{0.20\textwidth}
        \includegraphics[height=1\textwidth]{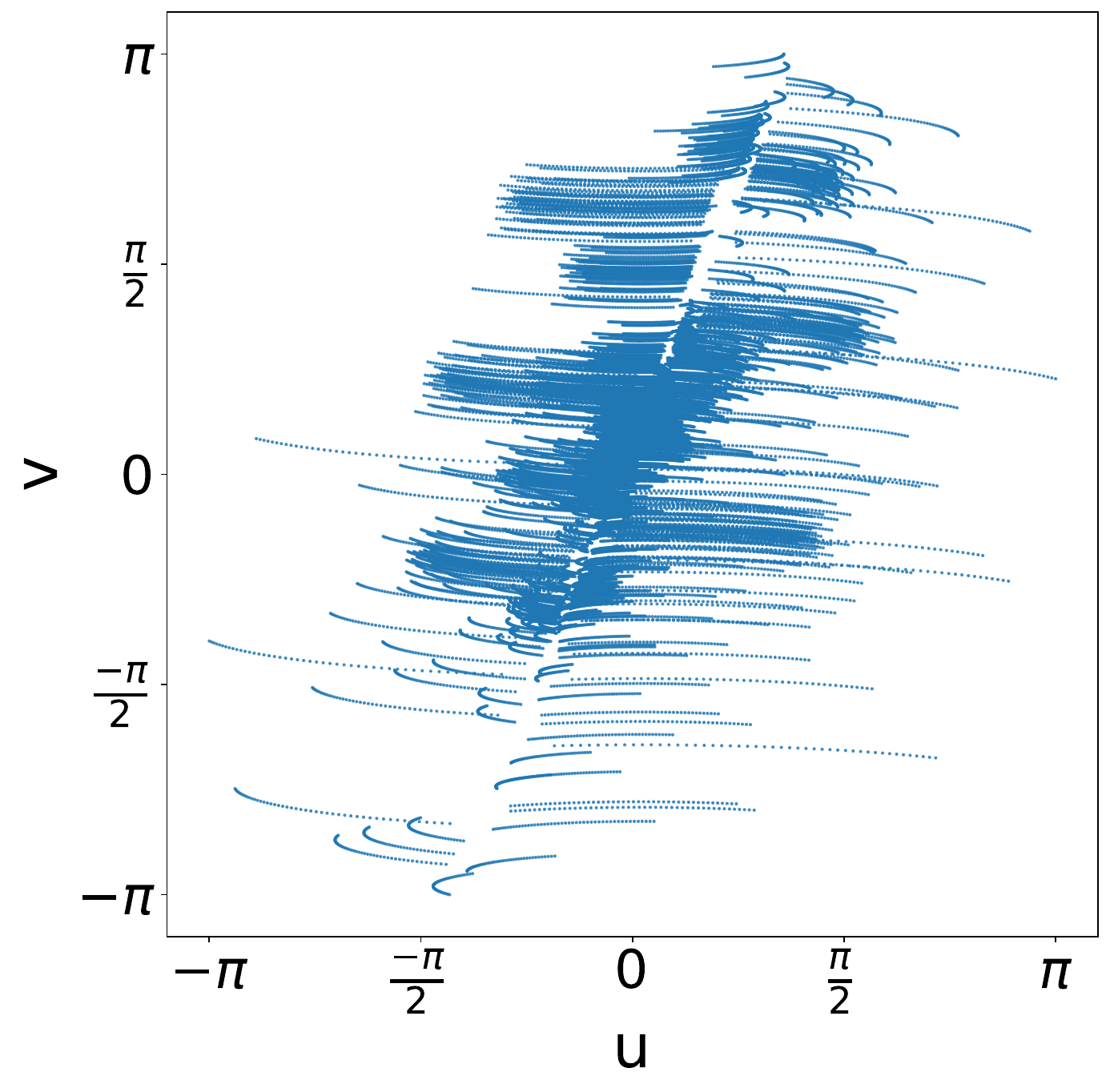}
    \end{subfigure}%
    \hspace{10pt}%
    \begin{subfigure}[t]{0.20\textwidth}
        \includegraphics[height=1\textwidth]{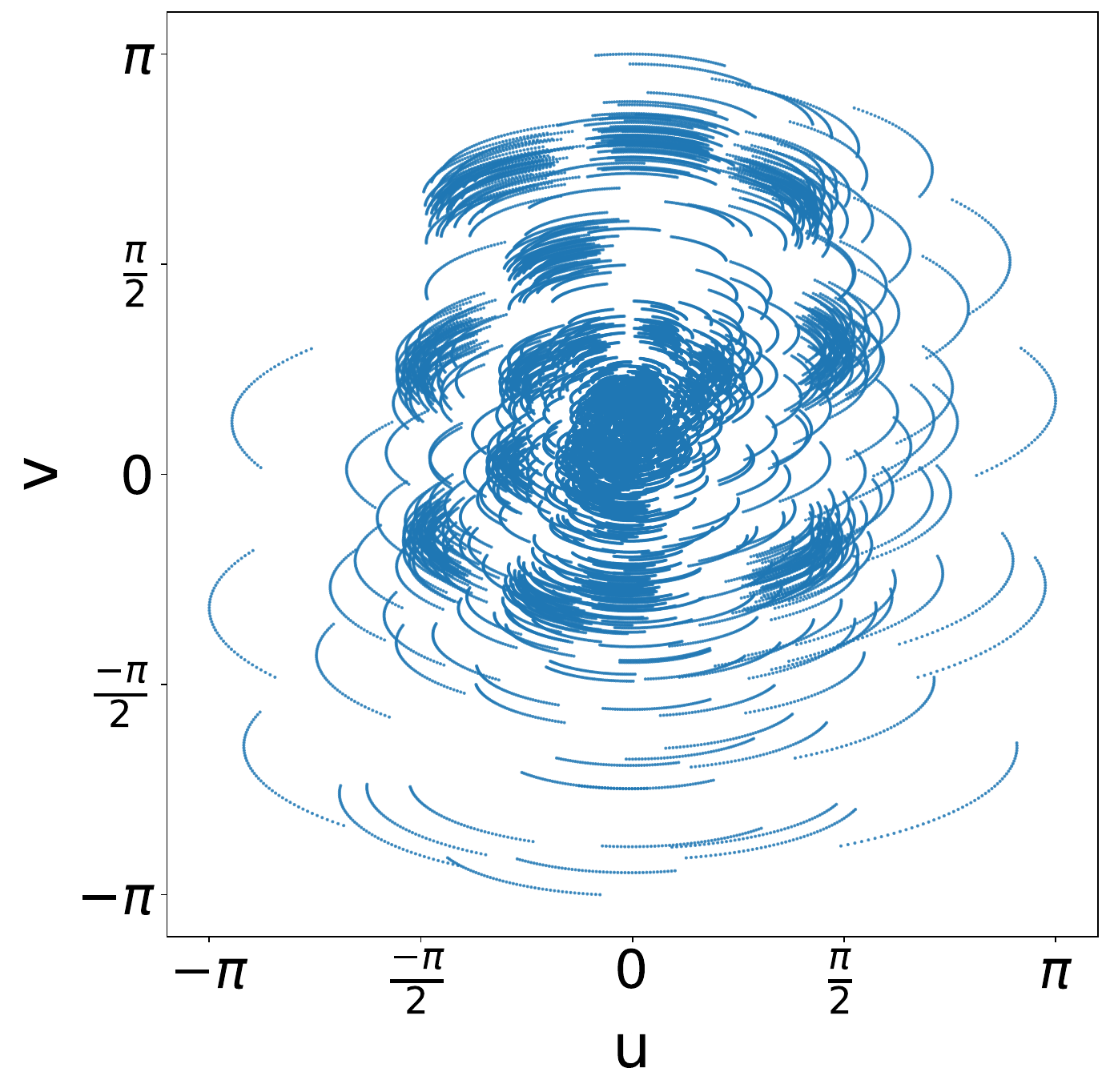}
    \end{subfigure}
    
    \caption{Examples of the simulated MeerKAT PSFs and their associated uv-coverage.}
    \label{fig:uv}
\end{figure}

The uv-coverages used are based on the MeerKAT \cite{Meerkat} observatory antenna positions. To simulate the PSFs, we use code from the \texttt{nenupy} \cite{nenupy} package. We present examples of uv-coverages and their associated PSFs in Figure \ref{fig:uv}.

\section{Conformalization}
\label{app:conf}

We suppose we have access to a reconstruction function $\hat{x}(\cdot)$ that maps an observed image $y \in \mathbb{R}^{d \times d}$ and a PSF $M \in \mathbb{R}^{d \times d}$ to a reconstructed image $\hat{x}(y, M)$. Our task is to create uncertainty intervals around each reconstructed image pixel. These intervals should contain the true pixel values with a user-specified probability. Formally, we construct, for each pixel $(m, n)$, the interval
\begin{equation}
    \mathcal{I}_{(m, n)}(y, M) = \left[ \hat{x}(y, M)_{(m, n)} - \Tilde{l}_{(m, n)}(y, M), \hat{x}(y, M)_{(m, n)} + \Tilde{u}_{(m, n)}(y, M) \right] \, ,
\end{equation}
where $\Tilde{l}$ and $\Tilde{u}$ are heuristic lower and upper interval lengths. These heuristics can, for example, be pixel-wise Gaussian standard deviations \cite{Nix1994}, softmax output distributions \cite{Rottmann2019}, pixel-wise quantiles \cite{Koenker1978}, or residual magnitude regressions. In this work, we assume $\Tilde{l} = \Tilde{u}$, and that they are both equal to the pixel-wise $90$-th quantile of the sampled bootstrap absolute residuals (see Section~\ref{sec:uq_methods}). The uncalibrated intervals are heuristic because they do not contain the ground truth with the desired probability, as we made no assumptions about the algorithm used to train $\Tilde{l}$ and $\Tilde{u}$.

In Risk-Controlling Prediction Set (RCPS) \cite{Angelopoulos2022}, the objective is to calibrate the interval $\mathcal{I}_{(m, n)}(y, M)$ such that
\begin{equation}
    P( \mathbb{E}[ \mathcal{S}(X, \, {\mathcal{I}}_{(m, n)}(y,M)) ] > \alpha ) \leq \delta  ,
    \label{eq:rcps}
\end{equation}
where $\mathcal{S}$ is an inteval score defined as
\begin{equation}
    \mathcal{S}(X_i, {\mathcal{I}}_{(m, n)}(y_i, M_i)) = 1 - \frac{| \{(m, n) : {X_i}_{(m, n)} \in {\mathcal{I}}_{(m, n)}(y_i, M_i) \} |}{d*d} ,
    \label{eq:rcps_score}
\end{equation}
which is the proportion of pixels where the ground truth is contained in the constructed intervals.
The inner expectation in Equation~\ref{eq:rcps} is over a new test point, while the outer probability is over the calibration set. Once calibrated, the intervals $\mathcal{I}_{(m, n)}(y, M)$ will an $(\alpha,\delta)$-RCPS. Let us suppose that we dispose of a calibration set ${(X_i, Y_i, M_i)}_{i \in \llbracket 1; N_C \rrbracket}$, where $N_C$ is the calibration set size and $X$ refers to the ground truth image we are hoping to reconstruct. Our goal is to calibrate the intervals using conformal prediction such that they become RCPS verifying Equation~\ref{eq:rcps}. We can obtain conformalized intervals as follows
\begin{equation}
\mathcal{I}^{(\lambda)}_{(m, n)}(y, M) = \left[ \hat{x}(y, M)_{(m, n)} - \lambda * \Tilde{l}_{(m, n)}(y, M), \hat{x}(y, M)_{(m, n)} + \lambda * \Tilde{u}_{(m, n)}(y, M) \right] ,
\end{equation}
where $\lambda$ should be chosen such that if $\alpha \in \left[ 0 , 1 \right]$ is a user-selected risk level and $\delta \in \left[ 0 , 1 \right]$ is a user-selected error level, then, the new intervals contain at least $1 - \alpha$ of the ground truth pixel values with probability $1 - \delta$. If we set $\alpha = \delta = 0.1$, then at least $90 \%$ of the ground truth images should have at least $90 \%$ of their pixels within the constructed intervals.

In practice, we take $N_C = 1000$ images with their corresponding PSFs. We then determine the heuristic intervals $\mathcal{I}^{(\lambda)}_i = \mathcal{I}_{(m,n)}^{(\lambda)}(y_i, M_i)$ for each element of this set. Later, we compute a score for each of these intervals $\mathcal{S}_i(\lambda) = \mathcal{S}(X_i, \mathcal{I}^{(\lambda)}_i)$ using Equation~\ref{eq:rcps_score}. In order to pick the smallest $\hat{\lambda}$ satisfying our goal, we follow \cite{Angelopoulos2022} and form Hoeffding's upper-confidence bound
\begin{equation}
    H(\lambda) = \frac{1}{N_C} \sum_{i = 1}^{N_C} \mathcal{S}_i(\lambda) + \sqrt{\frac{1}{2N_C} \text{log}\left(\frac{1}{\delta}\right)} .
\end{equation}
We can then pick $\hat{\lambda} = \{ \min_{\lambda} : H(\lambda) > \alpha \}$.

\section{Choice of the number of bootstrap samples}
\label{numberOfSamples}

\begin{figure}
    \centering
    \includegraphics[width=0.7\linewidth]{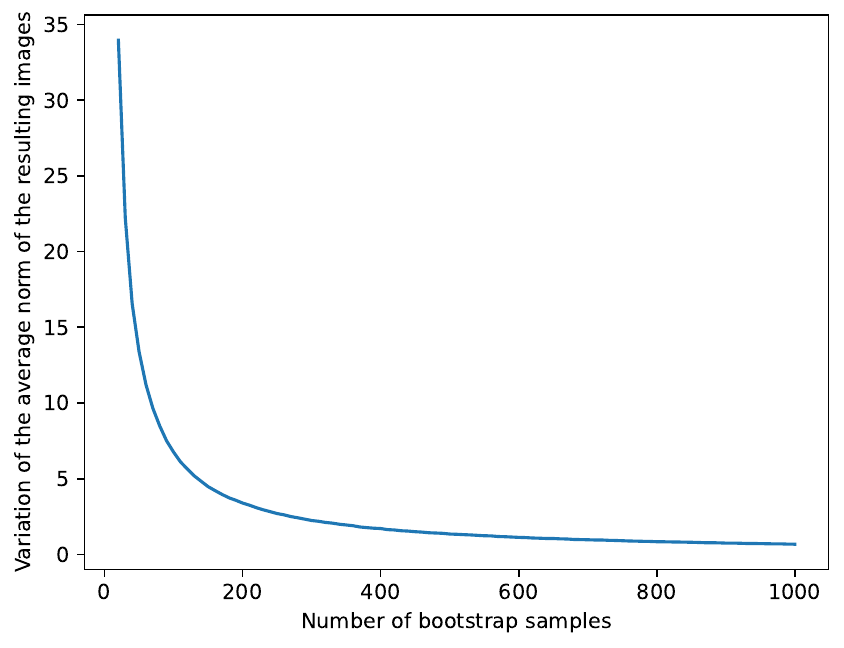}
    \caption{Variation of the $90$-th quantile estimation in CARB when increasing the number of bootstrap samples.}
    \label{fig:number_samples}
\end{figure}

We must choose the number of samples to use in the bootstrapping techniques. To accomplish this, we study the average variation of the norms of pixel-wise uncertainty images as we increase the number of samples. In practice, we use the $90$-th quantile error estimations described in Section~\ref{sec:uq_methods}. Figure~\ref{fig:number_samples} presents the curve obtained. We select a number of samples such that the curve is close to stagnating, which means that adding more samples to the bootstrap will not significantly change our confidence regions. This choice allows us to ensure sufficient samples while minimizing the run time of our algorithm.

\section{Additional results}
\label{additional_results}

\begin{figure}[h]

    \centering
    \begin{subfigure}[t]{0.12\textwidth}
        \includegraphics[height=1\textwidth]{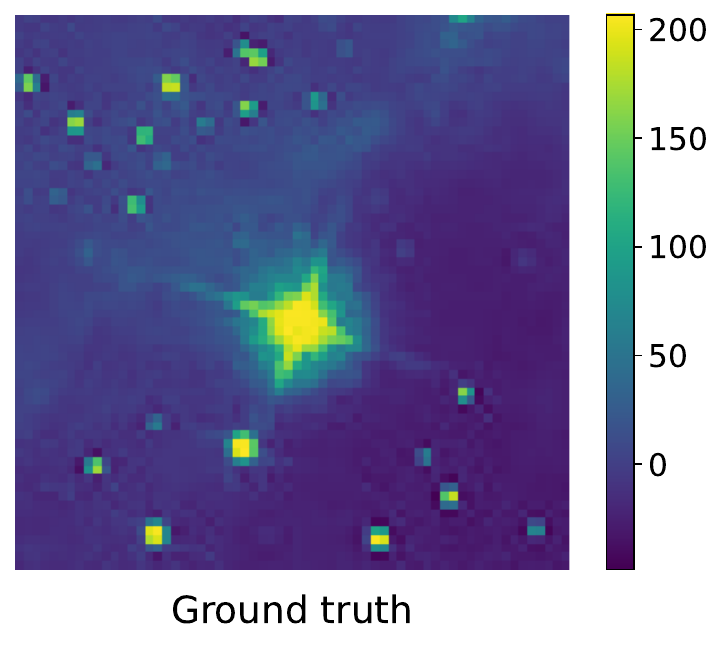}
    \end{subfigure}%
    \hspace{5pt}%
    \begin{subfigure}[t]{0.12\textwidth}
        \includegraphics[height=1\textwidth]{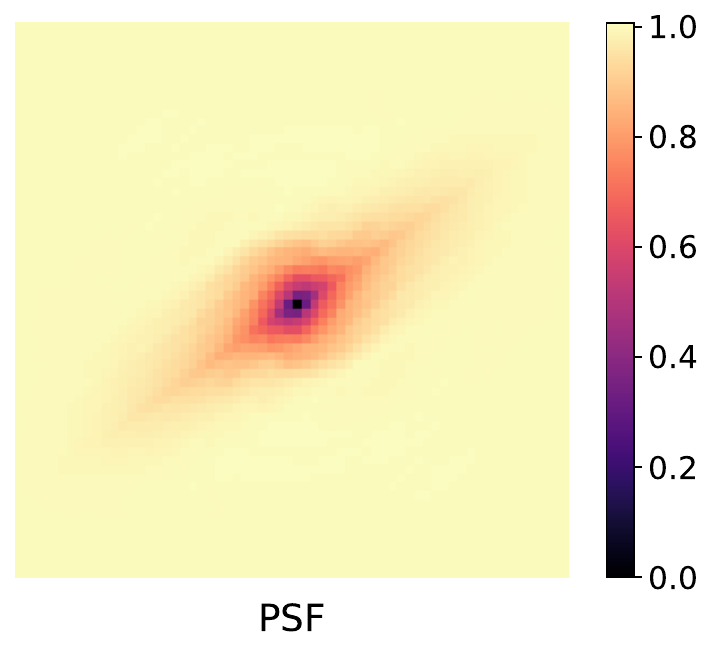}
    \end{subfigure}%
    \hspace{5pt}%
    \begin{subfigure}[t]{0.12\textwidth}
        \includegraphics[height=1\textwidth]{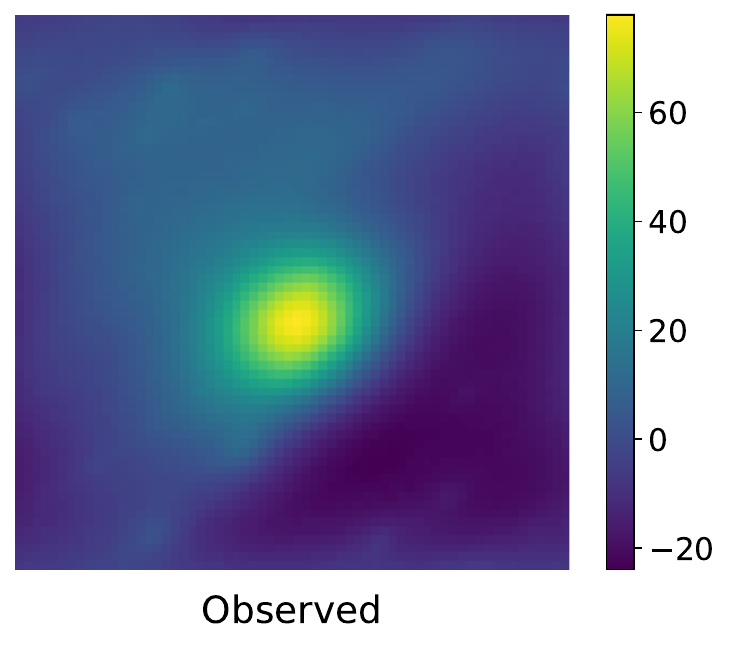}
    \end{subfigure}%
    \hspace{5pt}%
    \begin{subfigure}[t]{0.12\textwidth}
        \includegraphics[height=1\textwidth]{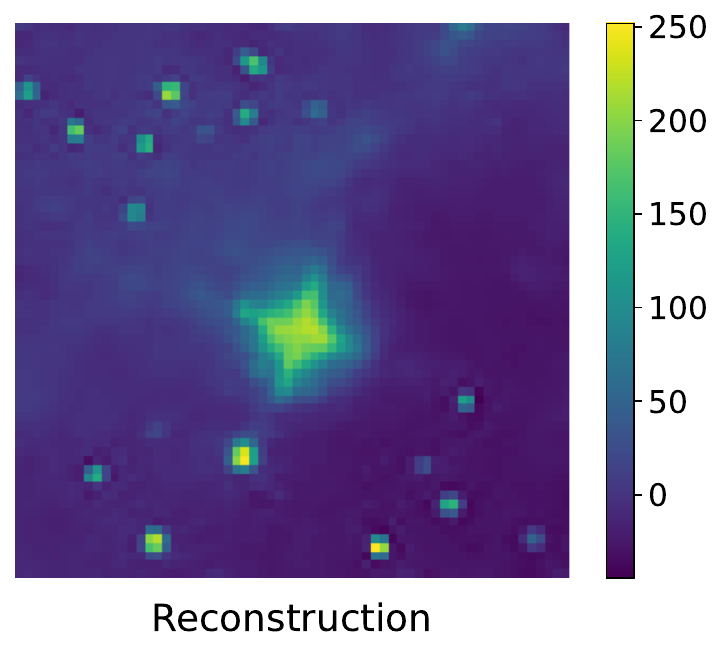}
    \end{subfigure}%
    \hspace{5pt}%
    \begin{subfigure}[t]{0.12\textwidth}
        \includegraphics[height=1\textwidth]{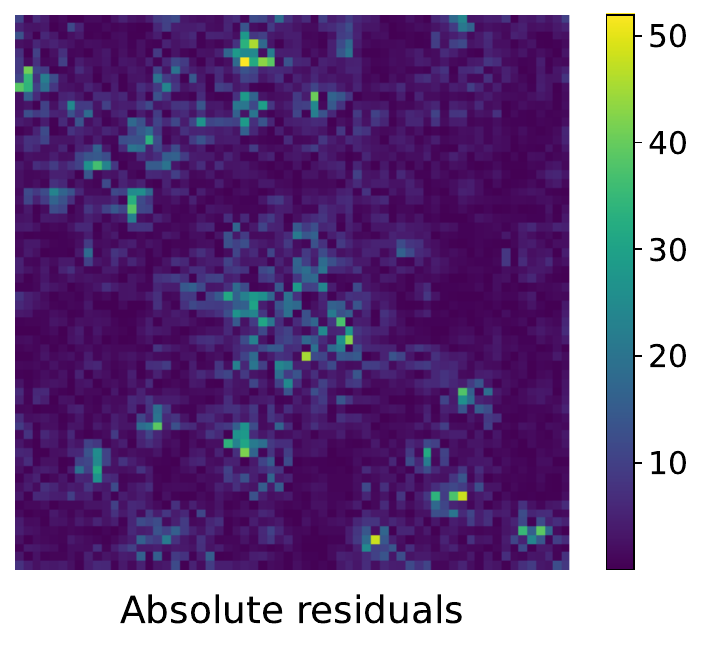}
    \end{subfigure}%
    \hspace{5pt}%
    \begin{subfigure}[t]{0.12\textwidth}
        \includegraphics[height=1\textwidth]{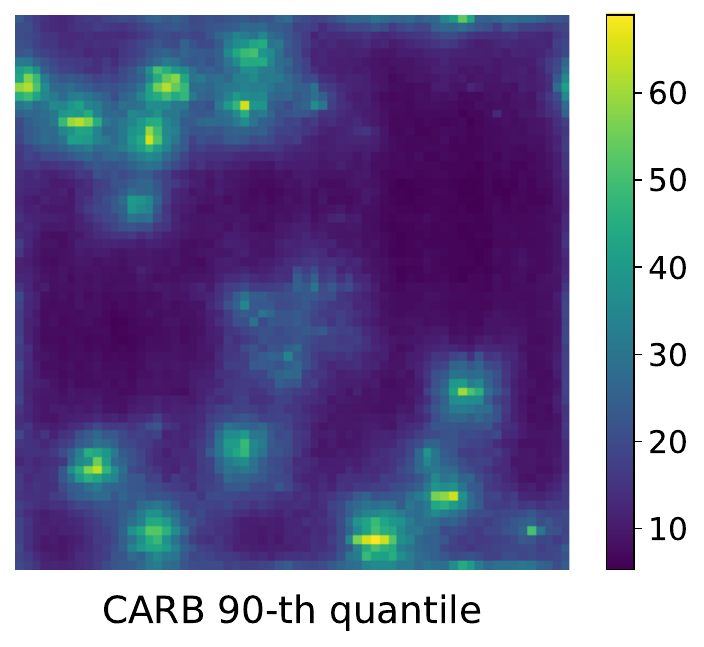}
    \end{subfigure}%
    \hspace{5pt}%
    \begin{subfigure}[t]{0.12\textwidth}
        \includegraphics[height=1\textwidth]{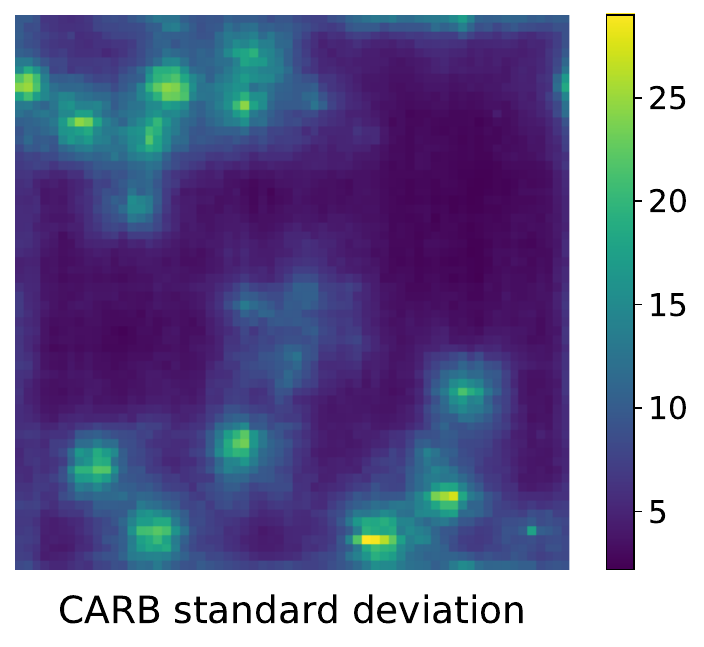}
    \end{subfigure}

    \centering
    \begin{subfigure}[t]{0.12\textwidth}
        \includegraphics[height=1\textwidth]{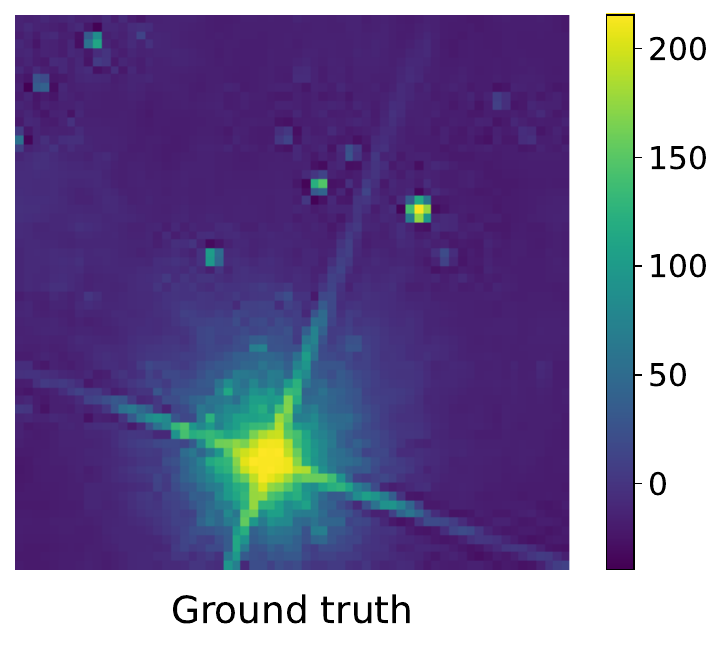}
    \end{subfigure}%
    \hspace{5pt}%
    \begin{subfigure}[t]{0.12\textwidth}
        \includegraphics[height=1\textwidth]{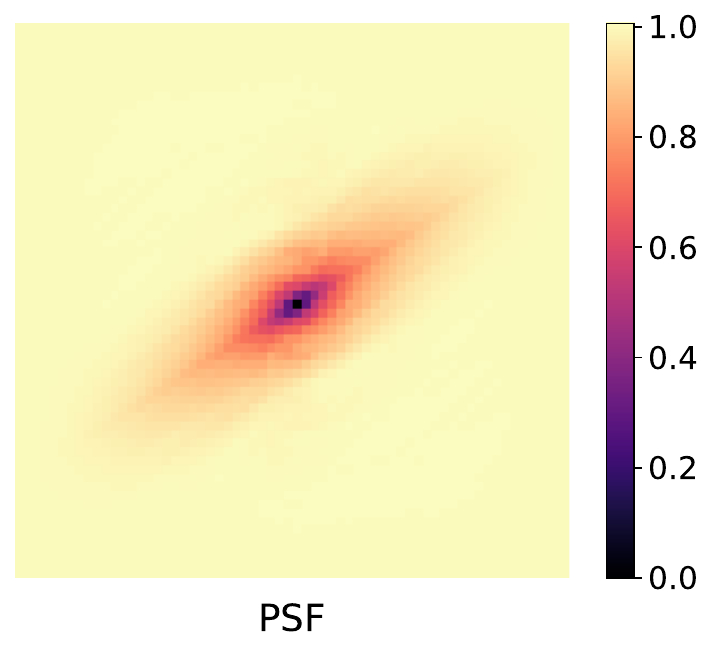}
    \end{subfigure}%
    \hspace{5pt}%
    \begin{subfigure}[t]{0.12\textwidth}
        \includegraphics[height=1\textwidth]{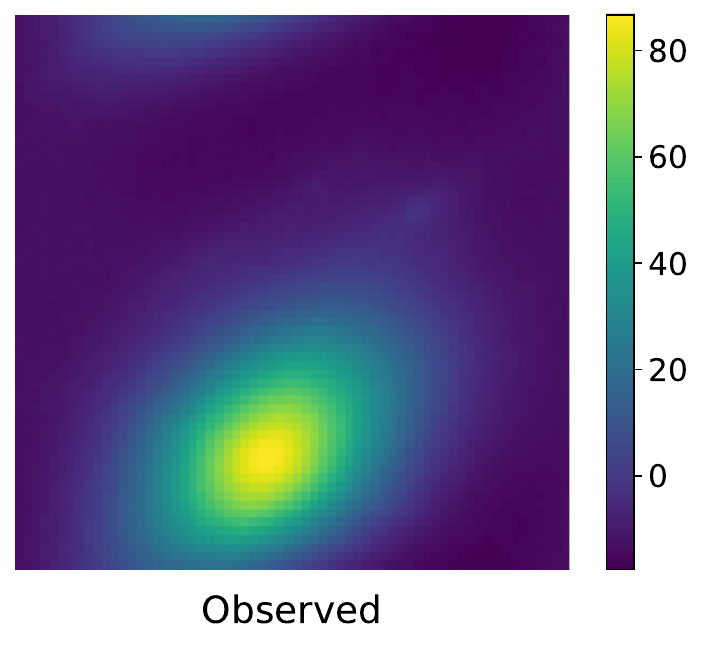}
    \end{subfigure}%
    \hspace{5pt}%
    \begin{subfigure}[t]{0.12\textwidth}
        \includegraphics[height=1\textwidth]{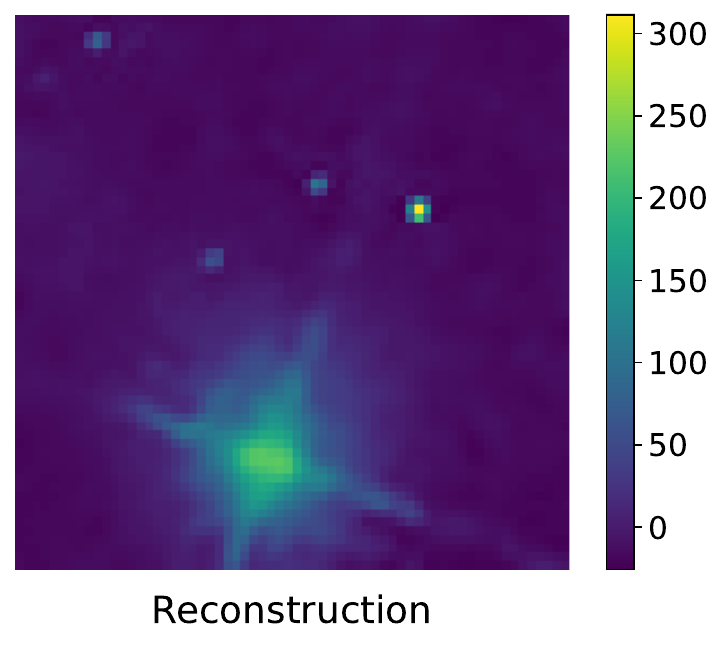}
    \end{subfigure}%
    \hspace{5pt}%
    \begin{subfigure}[t]{0.12\textwidth}
        \includegraphics[height=1\textwidth]{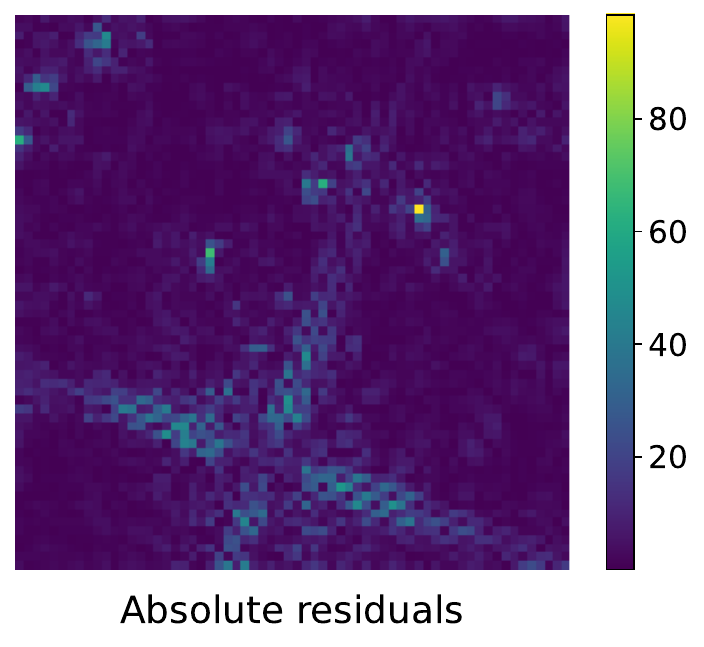}
    \end{subfigure}%
    \hspace{5pt}%
    \begin{subfigure}[t]{0.12\textwidth}
        \includegraphics[height=1\textwidth]{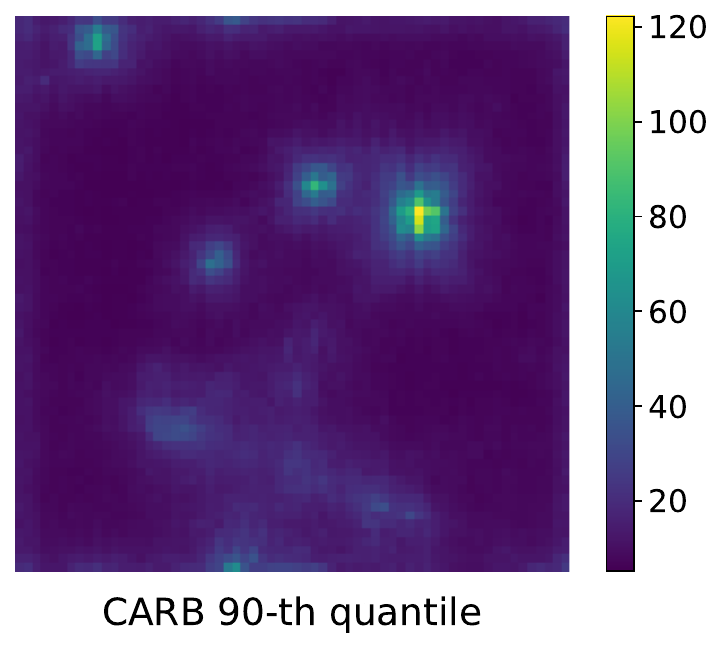}
    \end{subfigure}%
    \hspace{5pt}%
    \begin{subfigure}[t]{0.12\textwidth}
        \includegraphics[height=1\textwidth]{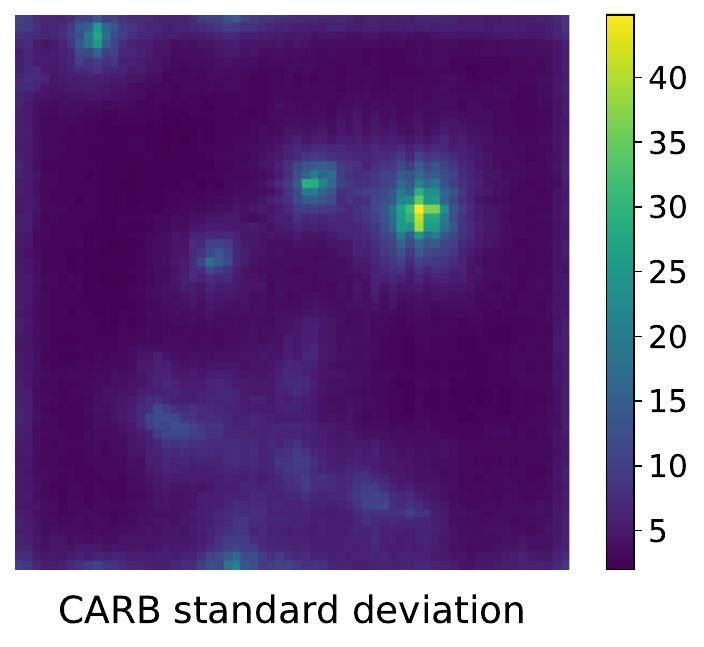}
    \end{subfigure}

    \centering
    \begin{subfigure}[t]{0.12\textwidth}
        \includegraphics[height=1\textwidth]{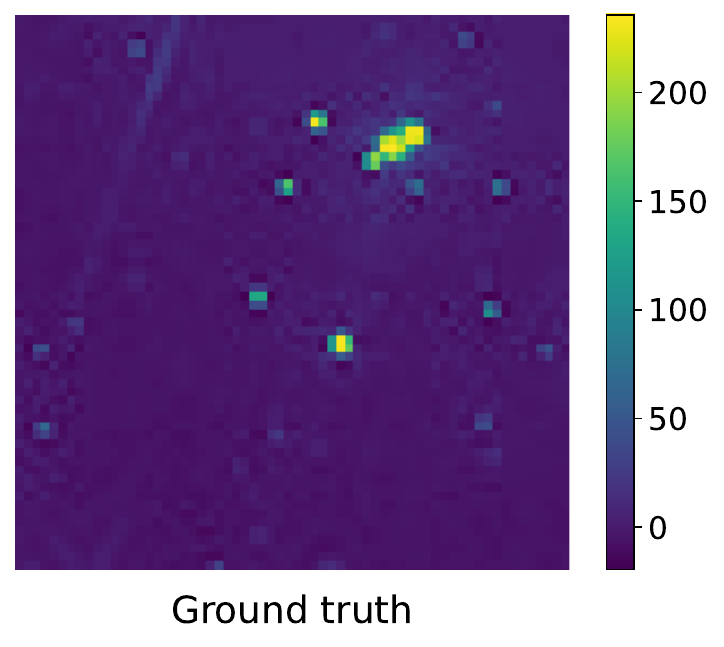}
    \end{subfigure}%
    \hspace{5pt}%
    \begin{subfigure}[t]{0.12\textwidth}
        \includegraphics[height=1\textwidth]{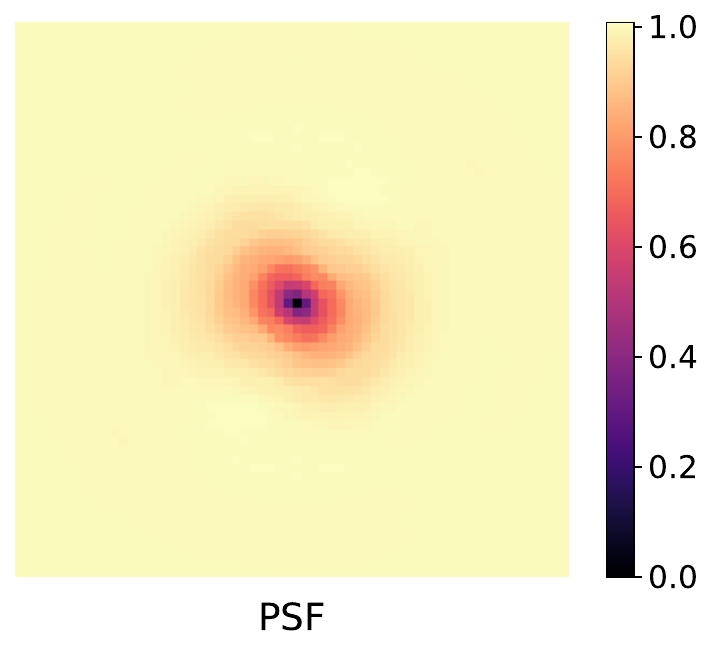}
    \end{subfigure}%
    \hspace{5pt}%
    \begin{subfigure}[t]{0.12\textwidth}
        \includegraphics[height=1\textwidth]{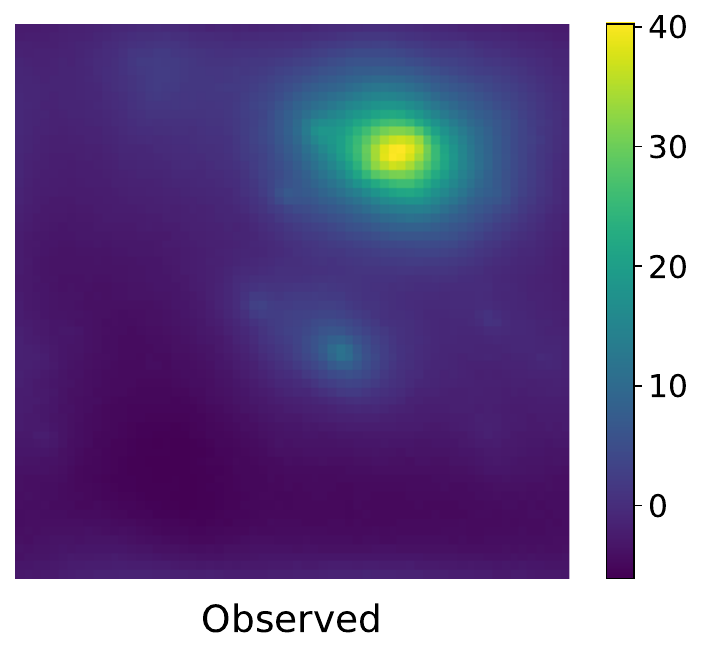}
    \end{subfigure}%
    \hspace{5pt}%
    \begin{subfigure}[t]{0.12\textwidth}
        \includegraphics[height=1\textwidth]{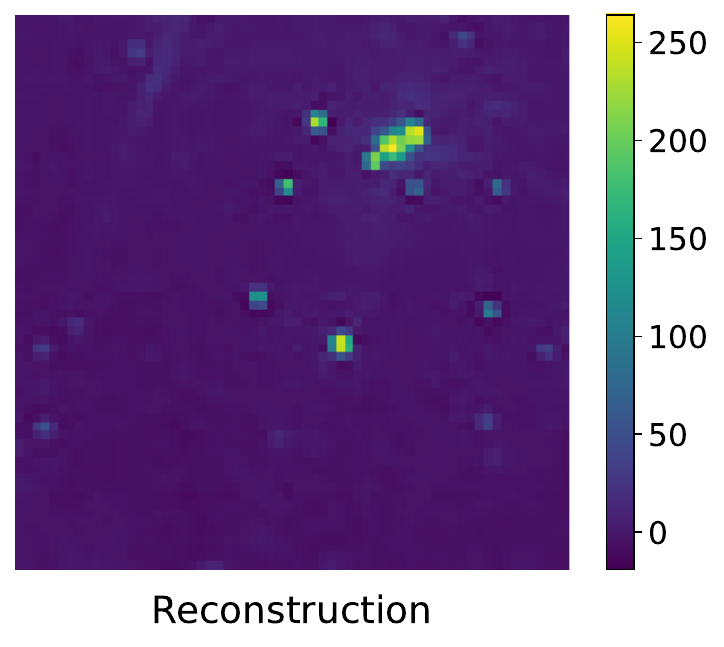}
    \end{subfigure}%
    \hspace{5pt}%
    \begin{subfigure}[t]{0.12\textwidth}
        \includegraphics[height=1\textwidth]{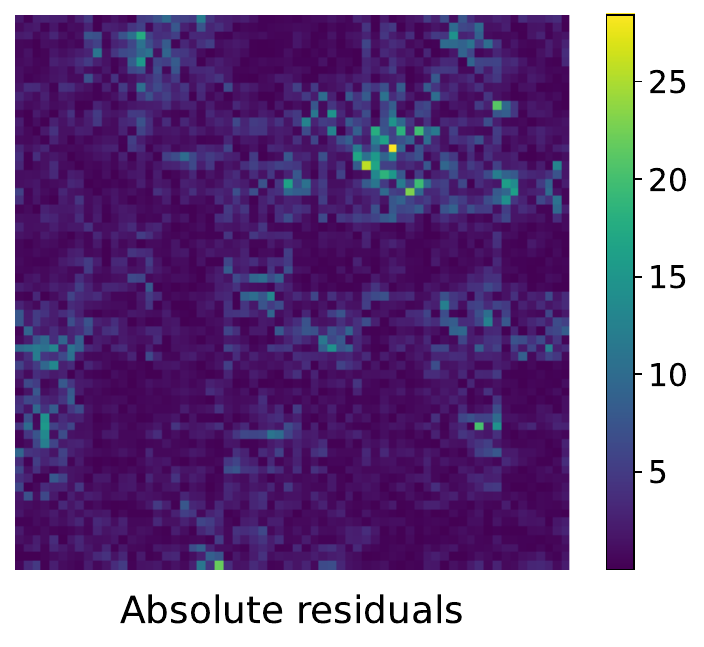}
    \end{subfigure}%
    \hspace{5pt}%
    \begin{subfigure}[t]{0.12\textwidth}
        \includegraphics[height=1\textwidth]{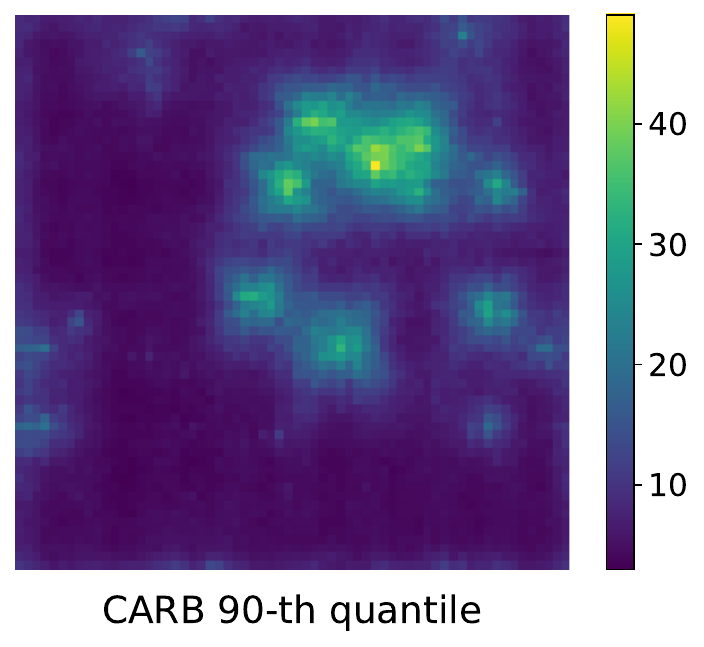}
    \end{subfigure}%
    \hspace{5pt}%
    \begin{subfigure}[t]{0.12\textwidth}
        \includegraphics[height=1\textwidth]{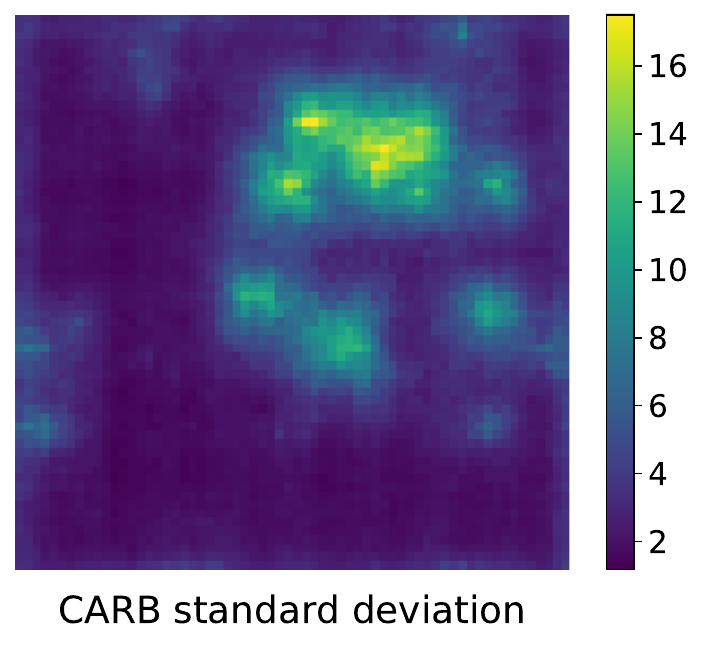}
    \end{subfigure}

    \centering
    \begin{subfigure}[t]{0.12\textwidth}
        \includegraphics[height=1\textwidth]{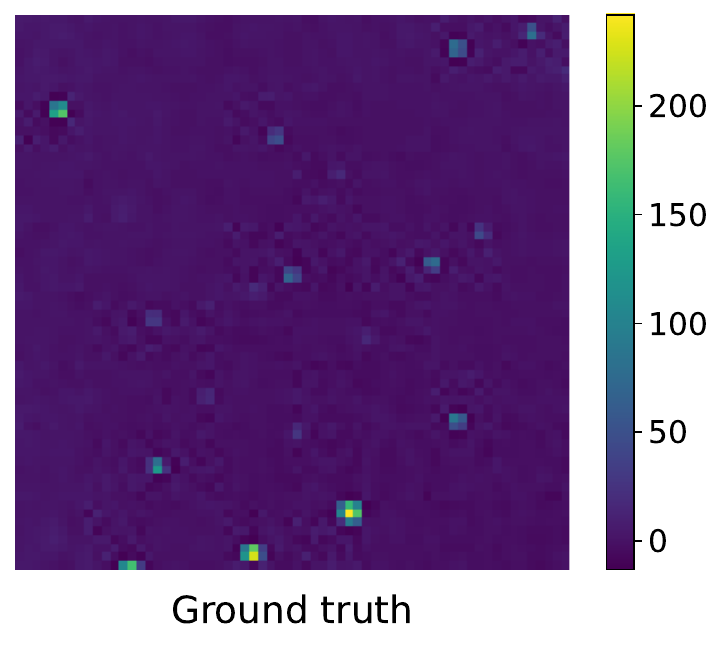}
    \end{subfigure}%
    \hspace{5pt}%
    \begin{subfigure}[t]{0.12\textwidth}
        \includegraphics[height=1\textwidth]{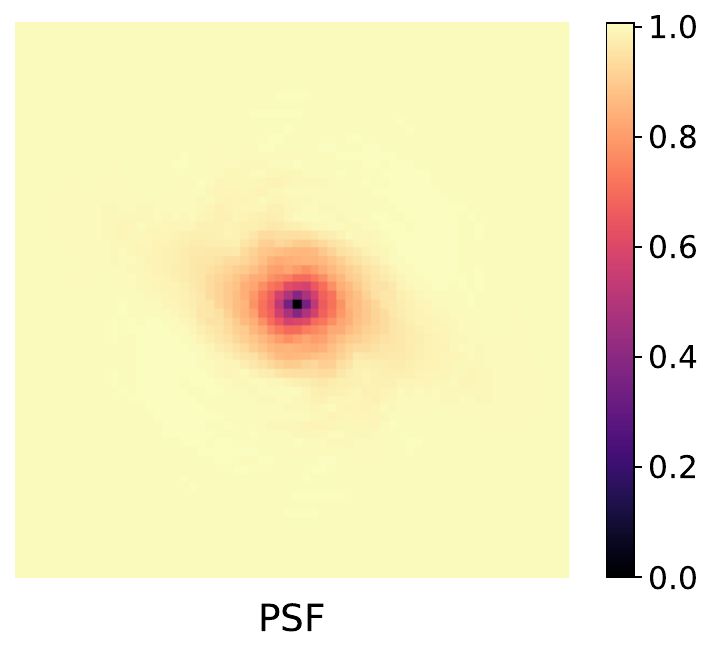}
    \end{subfigure}%
    \hspace{5pt}%
    \begin{subfigure}[t]{0.12\textwidth}
        \includegraphics[height=1\textwidth]{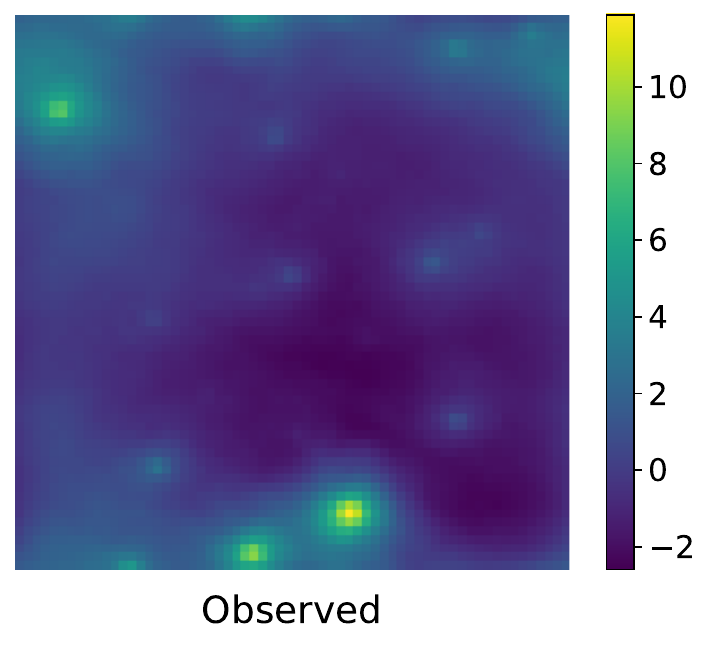}
    \end{subfigure}%
    \hspace{5pt}%
    \begin{subfigure}[t]{0.12\textwidth}
        \includegraphics[height=1\textwidth]{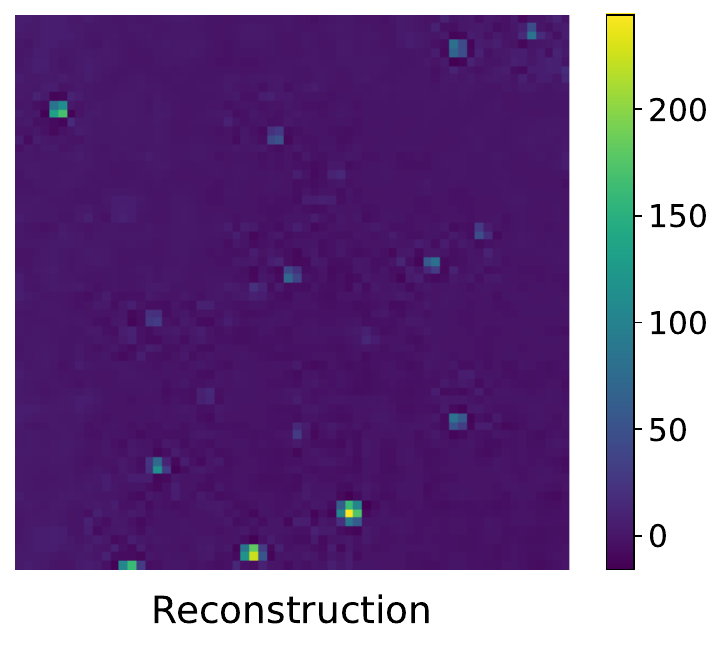}
    \end{subfigure}%
    \hspace{5pt}%
    \begin{subfigure}[t]{0.12\textwidth}
        \includegraphics[height=1\textwidth]{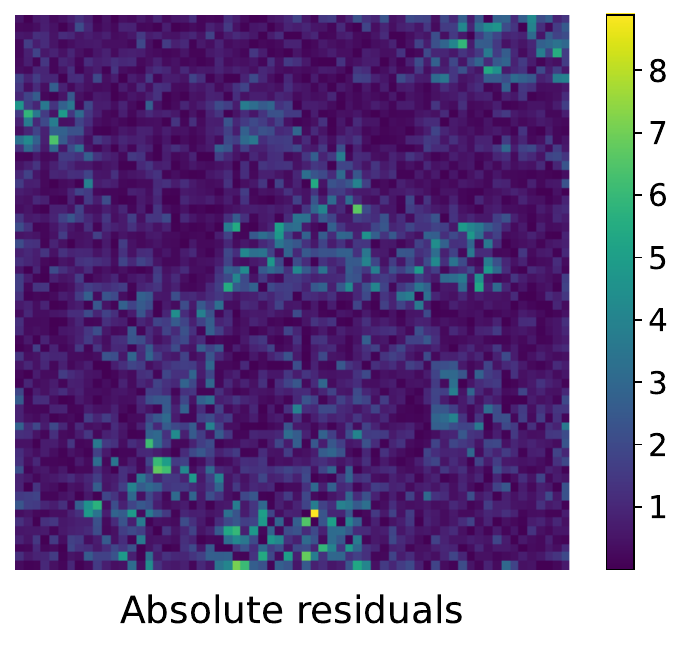}
    \end{subfigure}%
    \hspace{5pt}%
    \begin{subfigure}[t]{0.12\textwidth}
        \includegraphics[height=1\textwidth]{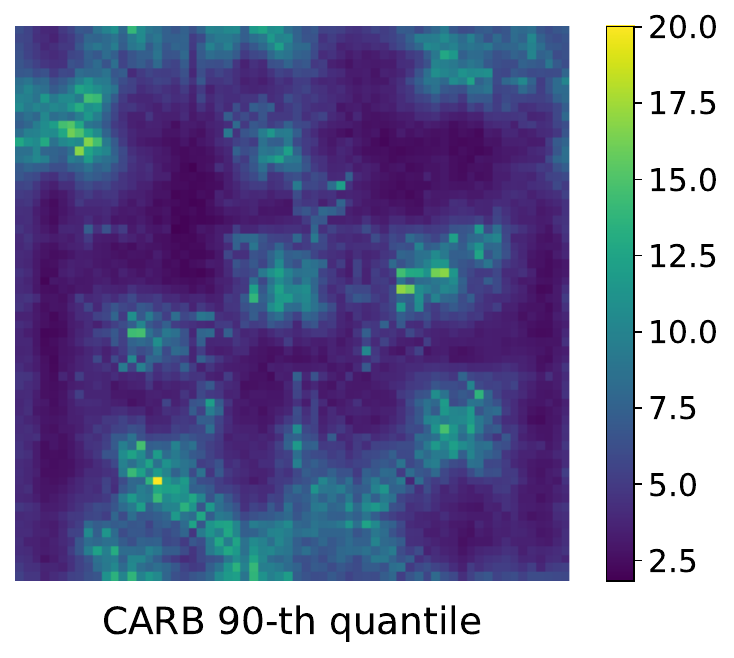}
    \end{subfigure}%
    \hspace{5pt}%
    \begin{subfigure}[t]{0.12\textwidth}
        \includegraphics[height=1\textwidth]{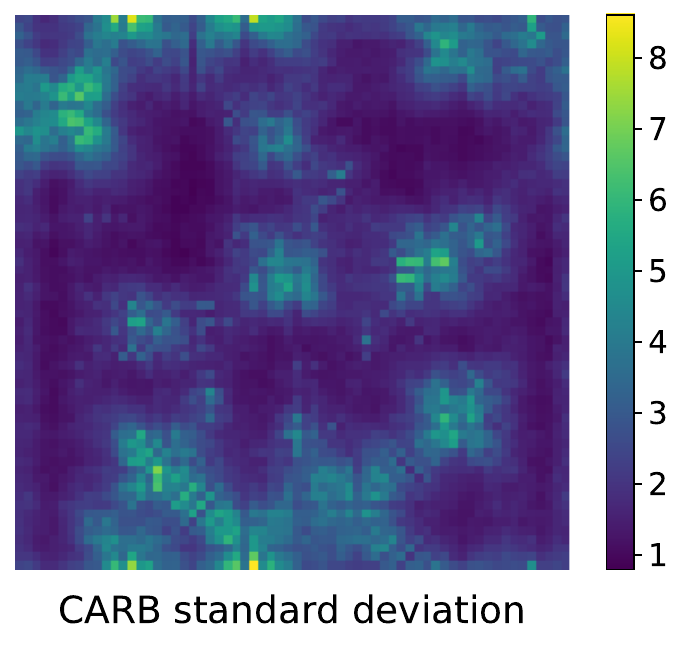}
    \end{subfigure}

    \centering
    \begin{subfigure}[t]{0.12\textwidth}
        \includegraphics[height=1\textwidth]{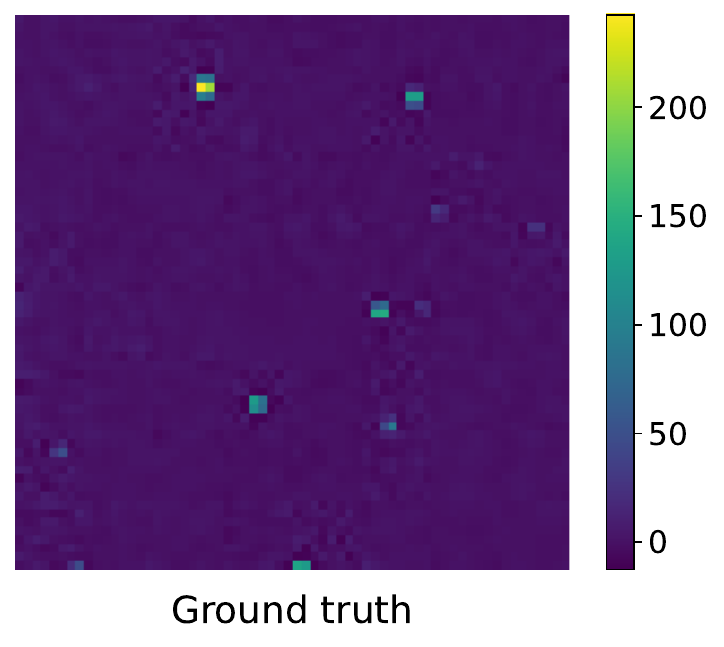}
    \end{subfigure}%
    \hspace{5pt}%
    \begin{subfigure}[t]{0.12\textwidth}
        \includegraphics[height=1\textwidth]{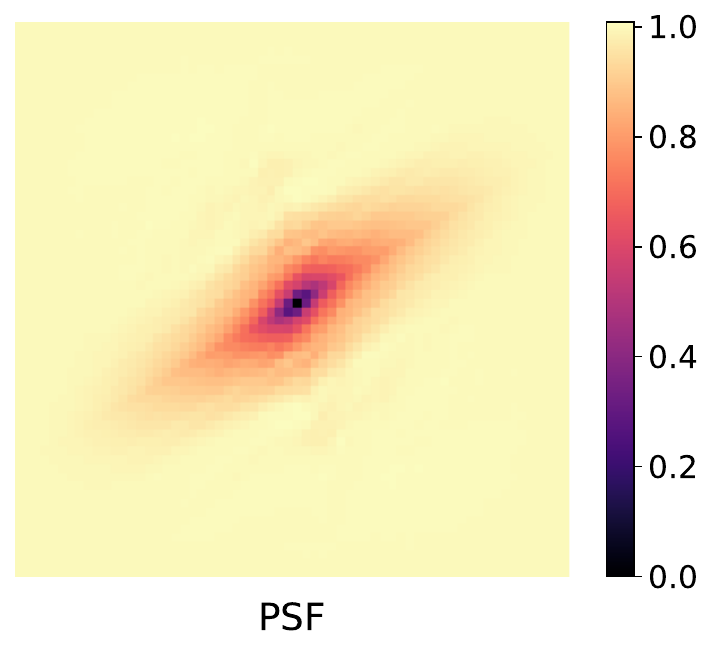}
    \end{subfigure}%
    \hspace{5pt}%
    \begin{subfigure}[t]{0.12\textwidth}
        \includegraphics[height=1\textwidth]{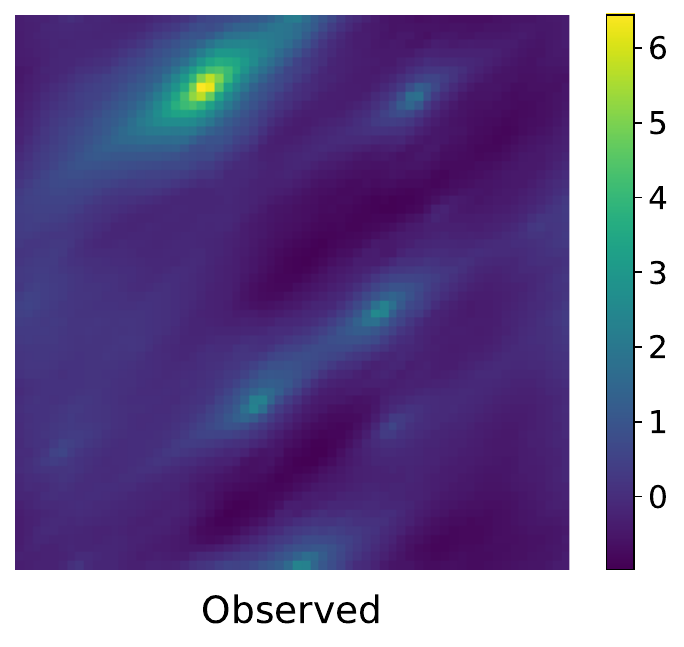}
    \end{subfigure}%
    \hspace{5pt}%
    \begin{subfigure}[t]{0.12\textwidth}
        \includegraphics[height=1\textwidth]{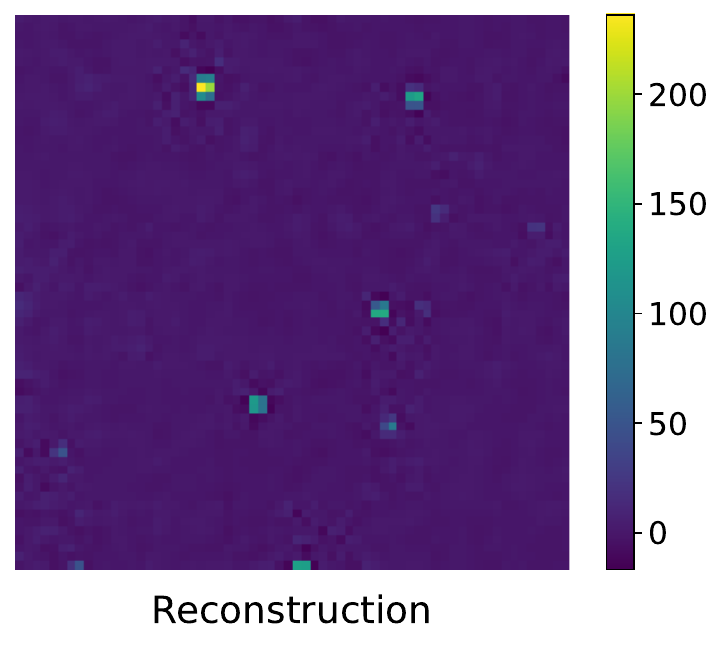}
    \end{subfigure}%
    \hspace{5pt}%
    \begin{subfigure}[t]{0.12\textwidth}
        \includegraphics[height=1\textwidth]{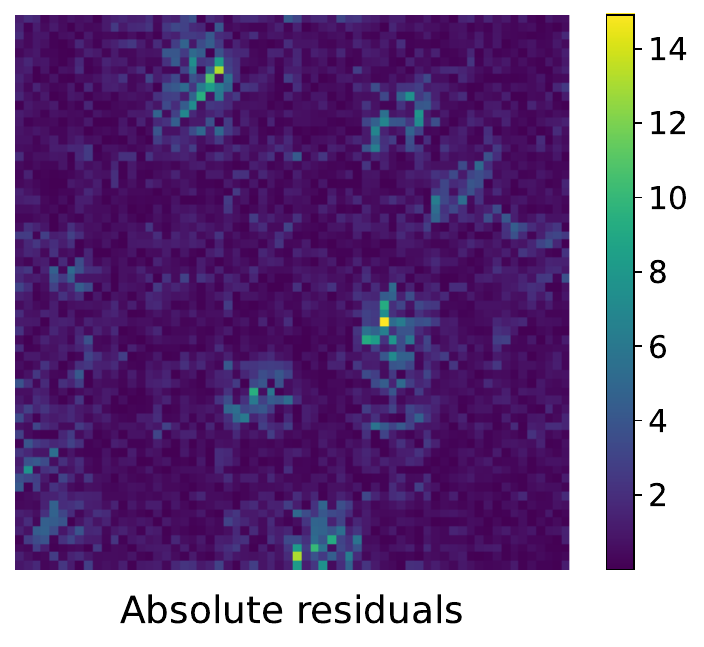}
    \end{subfigure}%
    \hspace{5pt}%
    \begin{subfigure}[t]{0.12\textwidth}
        \includegraphics[height=1\textwidth]{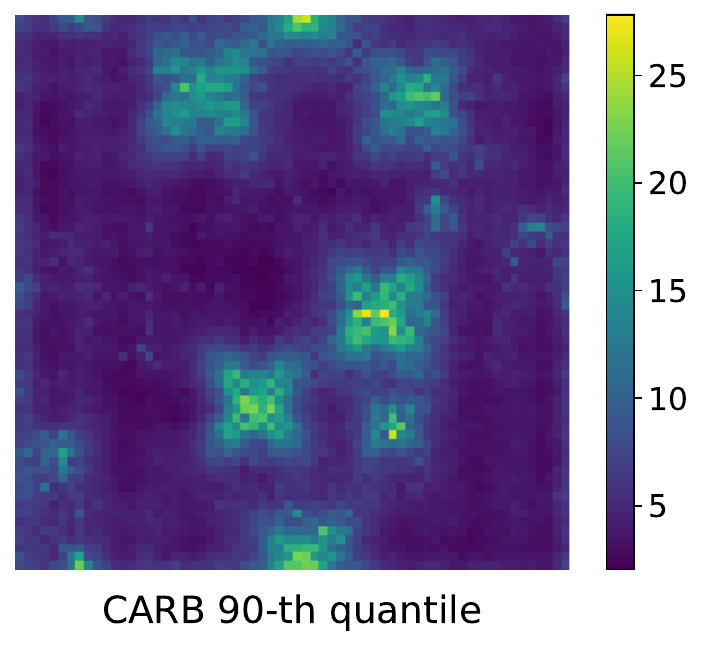}
    \end{subfigure}%
    \hspace{5pt}%
    \begin{subfigure}[t]{0.12\textwidth}
        \includegraphics[height=1\textwidth]{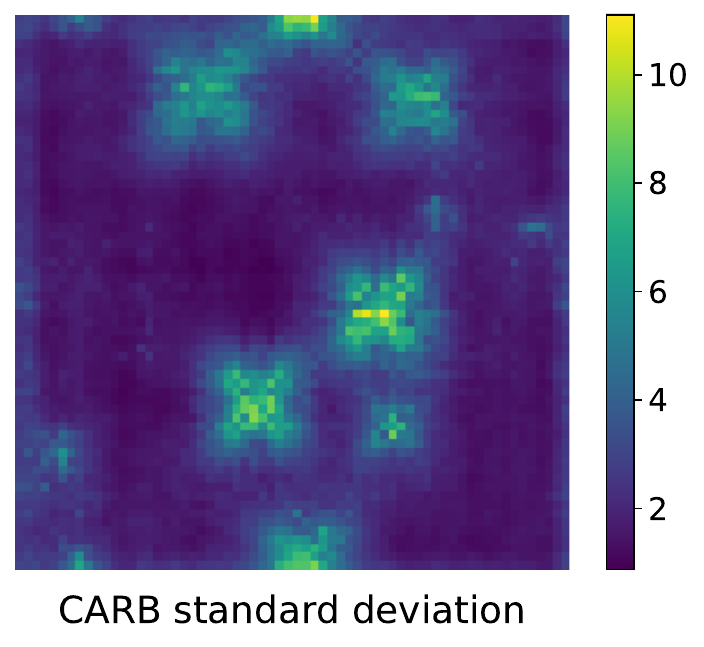}
    \end{subfigure}
    
    \caption{Additional radio interferometric image unrolling algorithm reconstruction results with uncertainty quantification from CARB. The first two columns show the ground truth images used with their associated PSF. The third column presents the observation, and the fourth column is the EVIL-Deconv reconstruction. The fifth column presents the oracle, or ground truth, absolute residuals. The last two columns introduce the pixel-wise uncertainty estimated from the CARB using the $90$-th quantile and the standard deviation of the bootstrap samples.}
    \label{fig:additional}
\end{figure}

Figure~\ref{fig:additional} shows additional astronomical images with their reconstructions and two methods to visualize the estimated CARB uncertainties. The first one takes the $90$-th quantile of each pixel of our sampled estimated errors, as shown in Figure~\ref{fig:error} for all bootstrap methods. The second alternative way of visualizing the pixel-wise uncertainties is to take the standard deviation of the computed bootstrap samples, as is done in \cite{Tachella2023}.

\end{document}